Development of an Autonomous Reverse Engineering Capability for Controller Area Network Messages to Support Autonomous Control Retrofits

Kevin Setterstrom
Department of Computer Science
North Dakota State University
1320 Albrecht Blvd., Room 258
Fargo, ND 58108

P: +1 (701) 231-8562
F: +1 (701) 231-8255
E: kevin.setterstrom@ndsu.edu

Jeremy Straub
Department of Computer Science
North Dakota State University
1320 Albrecht Blvd., Room 258
Fargo, ND 58108

P: +1 (701) 231-8196
F: +1 (701) 231-8255
E: jeremy.straub@ndsu.edu

**Abstract**

As the autonomous vehicle industry continues to grow, various companies are exploring the use of aftermarket kits to retrofit existing vehicles with semi-autonomous capabilities. However, differences in implementation of the controller area network (CAN) used by each vehicle manufacturer poses a significant challenge to achieving large-scale implementation of retrofits. To address this challenge, this research proposes a method for reverse engineering the CAN channels associated with a vehicle's accelerator and brake pedals, without any prior knowledge of the vehicle. By simultaneously recording inertial measurement unit (IMU) and CAN data during vehicle operation, the proposed algorithms can identify the CAN channels that correspond to each control. During testing of six vehicles from three manufacturers, the proposed method was shown to successfully identify the CAN channels for the accelerator pedal and brake pedal for each vehicle tested. These promising results demonstrate the potential for using this approach for developing aftermarket autonomous vehicle kits – potentially with additional research to facilitate real-time use. Notably, the proposed system has the potential to maintain its effectiveness despite changes in vehicle CAN standards, and it could potentially be adapted to function with any vehicle communications medium.

**1. Introduction**

The rapidly evolving landscape of autonomous vehicles presents the intriguing prospect of transforming conventional vehicles into self-driving ones. An alternative to purchasing an entirely new autonomous vehicle is the acquisition and installation of aftermarket kits, enabling the addition of semi-autonomous features at a comparatively affordable cost [1], [2]. However, a key challenge in the development of such kits is the variations across different vehicle models. One area of variation is the controller area network (CAN) that communicates different sensor data, topics, functions, and commands throughout vehicles' drive-by-wire architecture [3], [4]. The CAN system is a proprietary interface that varies between vehicle manufacturers [5]. There is no readily available documentation for interfacing with it, and each vehicle brand has a different CAN interface definition. This means that features developed for one specific vehicle may not work on another, as the CAN messages are different from vehicle to vehicle [6]. Companies wishing to provide aftermarket semi-autonomous upgrades are, thus, limited to supporting only the brands of cars that they can reverse engineer.

Despite the diverse CAN bus data found in different vehicle models, all modern automobiles share common user interfaces and serve similar purposes. Specifically, accelerator pedals are used to signal to propel the vehicle forward, while brake pedals are responsible for signaling deceleration. Given that input-corresponding messages are transmitted on the CAN bus [3], [4], it is reasonable to expect significant correlations between groups of CAN messages and specific vehicle responses. Even in the absence of prior knowledge about the CAN data, it is, thus, plausible to analyze the vehicle responses to deduce the meaning behind these messages.

This research employs an inertial measurement unit (IMU) to measure how a vehicle responds to operator inputs. By sampling vehicles' inertial data and CAN data simultaneously, the acceleration and deceleration CAN channels can be determined without any prior knowledge of the vehicle's CAN system. Based on the data and analysis presented, which was gathered from multiple vehicles, this method is shown to be feasible to use to reverse engineer the acceleration and deceleration CAN channels of any vehicle and to identify the key messages related to the vehicle's accelerator and brake pedals.

In addition to retrofitting vehicles with autonomous driving capabilities, this research also has potential benefits for vehicle threat analysis and cyber protection. With drive-by-wire technology present in modern vehicles, a system that can interpret and understand the specific CAN messages required for the vehicle to function could potentially use those messages to gain control of the vehicle. This capability can also monitor those messages to detect abnormalities that could indicate unauthorized communications within a vehicle.

## 2. Background

This section reviews the context for and foundational knowledge related to autonomous reverse engineering of automotive CAN. The evolution of the CAN bus and its significance in modern vehicles is explained. Then, the technical aspects of CAN, including its communication protocol and the structure of CAN frames, are discussed. Finally, existing methods and applications for automating the reverse engineering process and literature that highlights the vulnerability of vehicle CAN systems to attacks are reviewed.

The subsequent sections of this paper build upon this knowledge, and present a research methodology and findings related to the autonomous reverse engineering of acceleration and deceleration CAN channels.

### *2.1. Controller Area Network Bus*

In the early 1980's, Bosch introduced the CAN bus serial communication protocol to help address the increasing vehicle emissions reduction requirements and reduce the wire count in the automotive industry [7]. Since its inception, the CAN bus has become the de facto communication standard across the automotive industry, with Mercedes becoming the first automotive manufacturer to implement the protocol [8]. The introduction of the CAN bus brought about significant changes in basic system monitoring and vehicle control [3], [4], [9]. One notable change has been the transformation of how vehicles process inputs from the operator, particularly in relation to engine operation and vehicle movement. In spark ignition engines, the power generated by the engine is directly influenced by the amount of air entering the intake manifold. Traditionally, the angular position of the throttle valve, which regulates the airflow, was controlled by a mechanical link connected to the accelerator pedal, operated by the driver [4]. This mechanical linkage physically translated the driver's input into an adjustment of the throttle valve, which is now done electronically.

The increase in requirements relating to vehicle controls, drivability, and safety have led to the development of drive by wire technology [3], [4]. Drive-by-wire refers to the integration of the CAN bus into the vehicle's control system. With drive-by-wire technology, the mechanical link between the accelerator pedal and the throttle valve has been eliminated. Instead, sensors, actuators, and electronic control units (ECUs) communicate over the CAN bus to command these functions. For instance, an electronic system known as a motorized throttle body now controls the amount of air entering the intake manifold, and sensors measure the angle of the accelerator pedal. When the operator presses the accelerator pedal, it produces an electrical input that is interpreted by an ECU and transmitted through the vehicle's CAN system. Further along the CAN bus, another ECU interprets the CAN message, extracts the throttle position sensor data, and electrically adjusts the vehicle's intake. This digitalization of the control system, made possible by the CAN bus, produces the same vehicle response as the previous mechanical linkage technology, but with enhanced precision and efficiency. A similar concept is also applied to the braking functions within the vehicle.

### *2.2. Controller Area Network Implementation*

Modern vehicles continue to include an increasing number of electronic components [3], [23], [24]. A majority of these electronics are referred to as ECUs, which are small, embedded computers. ECUs utilize sensors and electromechanical actuators to monitor and control various vehicle features [23]. The introduction of ECUs has provided improved on-board diagnostic capabilities, while also paving a way for enhanced driver assistance functionality and vehicle autonomy. Vehicle ECUs communicate with one another using serial communication protocols [25].

One of the most common standards that is used is CAN [26]. CAN is an International Standardization Organization (ISO) defined serial communication bus that was originally developed for the automotive industry to replace the complex wiring harnesses used by the two-wire bus architecture [27]. Since then, it has been adopted by many other industries such as marine, medical, aerospace, and manufacturing [8]. Figure 1 shows an example of a serial network implemented without CAN (left) and contrasts that to a network implemented with CAN (right). The left side of the image illustrates the complexity and the larger number of wires required in a network without CAN bus. The right side demonstrates the simplicity and reduced wire count necessary to achieve the same system functions by utilizing the CAN bus.

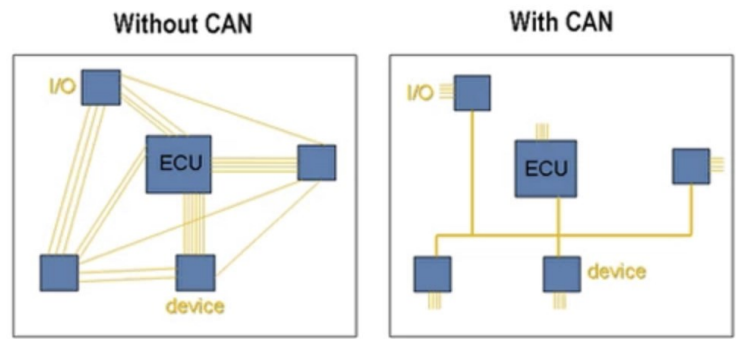

**Figure 1.** Automotive control network without CAN (left) vs with CAN (right) [28].

The CAN communications protocol, ISO 11898, describes how information is passed between devices within the network. ISO 11898 conforms to the Open Systems Interconnection (OSI) [29] model that is defined in terms of layers. The ISO 11898-1 [30] architecture implements the lowest two layers of the seven-layer OSI/ISO model, the data-link layer and the physical layer. The data-link layer's role is to transmit messages from a node to the network reliably, ensuring error-free communication. It manages tasks such as bit stuffing and checksum calculations. After sending a message, it awaits acknowledgement from the receivers [31]. The physical layer is the hardware infrastructure of a CAN network, encompassing the ISO 11898 electrical specifications. It facilitates the conversion of binary 1's and 0's into electrical pulses, when leaving a node, and vice versa when receiving CAN messages at a node [31]. Figure 2 illustrates the OSI layers used in CAN.

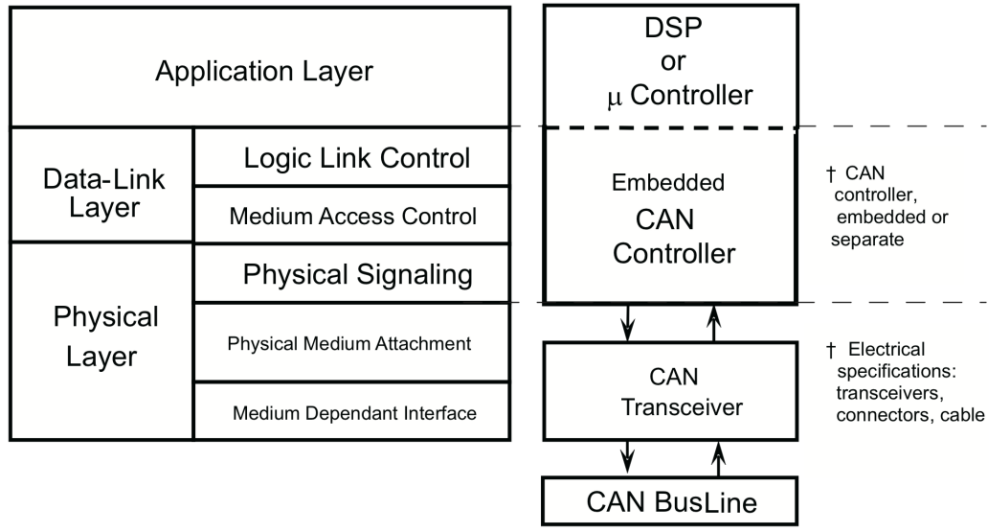

**Figure 2.** OSI layers used in CAN [31].

The data-link layer and physical layer define how controller area networks function. To communicate with an existing CAN bus, a user also needs to know details relating to its transmission frequencies. ISO 11898-2 [32] specifies the high-speed physical medium of the controller area network, which supports transmission rates up to 1Mbit/second. The high-speed CAN (HSC) for vehicle applications is further defined in the standard J2284/3_201611 and communicates at a rate of 500 kbps [33].

Next, the implementation of CAN at the application level is reviewed. CAN messages are formatted and structured into CAN frames. Figure 3 and Figure 4 show CAN frames as implemented by Texas instruments [34].

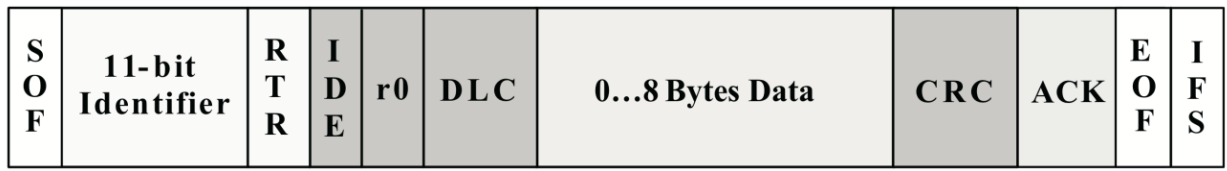

**Figure 3.** Standard CAN: 11-Bit Identifier [34].

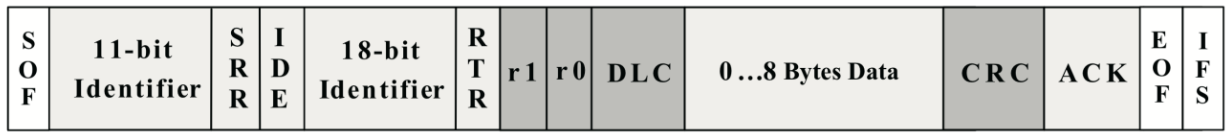

**Figure 4.** Extended CAN: 29-Bit Identifier [34].

Reverse engineering research focuses mostly on the identifier, data length code (DLC), and the data sections of the CAN Frame. The identifier is what determines the arbitration of the CAN message. Higher priority messages are indicated with lower binary values. The DLC indicates how many bytes the data field contains. As shown in Figure 3 and 4, the DLC can range from zero to eight bytes. The data field, also known as the CAN frame payload, includes information that can be used for different applications. For example, if eight bytes of payload information is present, it is likely that not all eight bytes of this data should be grouped together. Typically, it is grouped or segregated into channels.

The biggest difference between standard CAN (CAN 2.0A) and extended CAN (CAN 2.0B) is the number of possible ID's available in the CAN frame architecture. Extended CAN increased the number of possible messages from the standard CAN by using an 18-bit identifier.

Understanding how CAN frame data is divided into CAN channels is important for this research. Table 1 shows CAN frame information, including the arbitration ID (564), and 8 bytes (64 bits) worth of data. Not all 64 bits of data relate to the same feature. Instead, there are 8 different features that are communicated in this single CAN frame. The frame is comprised of 8 channels which are 8 bits (1 byte) each, as shown in Table 2. In this example, a specific CAN channel would be indicated as: CAN frame ID: 564, data: byte 0.

**Table 1.** Example CAN frame.

| CAN Frame | | | | | | | | |
|---|---|---|---|---|---|---|---|---|
| Arbitration ID | Data | | | | | | | |
| 564 | 64 bits | | | | | | | |
| | Byte_0 | Byte_1 | Byte_2 | Byte_3 | Byte_4 | Byte_5 | Byte_6 | Byte_7 |

**Table 2.** Example CAN channels.

| CAN Channels | | |
|---|---|---|
| Arbitration ID | Data | Channel ID |
| 564 | Byte_0 | 1 |
| 564 | Byte_1 | 2 |
| 564 | Byte_2 | 3 |
| 564 | Byte_3 | 4 |
| 564 | Byte_4 | 5 |
| 564 | Byte_5 | 6 |
| 564 | Byte_6 | 7 |
| 564 | Byte_7 | 8 |

The example shown in Tables 1 and 2 demonstrates how a CAN frame can be broken into channels that are 1 byte each. However, CAN frames are sometimes not grouped together evenly [5], [10], [20]. The CAN channel start and end bits can be any consecutive combination within the CAN frame. Marchetti and Stabili [10] refer to these groups (CAN channels) as "tokens" and the discovery of the boundaries of each token was referred to as "tokenization".

### 2.3. CAN Bus Reverse Engineering

The automotive industry is beginning to include self-driving capabilities in vehicles which are also growing progressively more connected.

Automating the process of reverse engineering vehicle CAN systems has many potential benefits. Several methods have been proposed to achieve this goal. Marchetti and Stabili's READ algorithm [10], developed in 2018, analyzes CAN traffic containing unknown messages to automatically identify and label distinct types of signals encoded in their data frames. The algorithm discovers signal boundaries without prior knowledge of the vehicle's CAN bus.

Huybrechts, *et al.* [18], proposed an automated reverse engineering system for automotive CAN data in the same year. They proposed two solutions: an arithmetic approach and a machine learning approach, which both compare known OBD-II PID signals with raw CAN messages on the vehicle's CAN system. Both methods showed promising results. Kang, *et al.* [19], also proposed a version of an automated automotive CAN reverse engineering system called automated CAN analyzer (ACA). This system uses a relational comparison approach between the response data of known diagnostic queries and CAN traffic data. They proposed an automated function that injects fake messages into the CAN bus, based on the pre-analyzed CAN messages discovered by their ACA system.

Pese, *et al.* developed LibreCAN [20] in 2019, which uses a three-phase approach: signal extraction, identifying kinematic-related data, and identifying body-related data. Their work draws inspiration from Marchetti and Stabili's READ algorithm [10] and used DBC files, OBD-II PID messages, and cross correlation. They also used a diagnostic app named "Torque Pro" [21] to provide inertial measurements to help identify accelerometer-related data within the CAN bus. CANMatch [5] is a recent improvement to LibreCAN that introduced frame matching to their equation and exploited the reuse of CAN frame IDs among vehicle models to achieve fully automated CAN bus reverse engineering. To aid in CAN frame matching, while also acting as a form of ground truth, Database CAN (DBC) files are used.

Blaauwedraad and Kieberl [22] demonstrated the automated reverse-engineering of CAN messages using OBD-II and correlation coefficients, which looked for a direct match between raw CAN data and specific OBD-II PID responses.

All of these approaches have shown promising benefits for reverse engineering vehicle CAN systems.

*2.4. Security Considerations*

While technological advancements aim to improve the driving experience and overall safety, they also introduce a larger attack surface that malicious actors can potentially exploit to compromise the security of connected vehicles [10]. Miller and Valasek [11]–[13] have demonstrated the vulnerability of vehicle CAN systems to attacks, including wireless threats. For instance, they showed how an unmodified 2014 Jeep Cherokee can be remotely attacked, enabling the attacker to take control of the vehicle [13]. Other similar examples include attacks against BMW's Connected Drive [14], GM's OnStar [15], and Tesla vehicles [16], [17].

## 3. Research Methodology

This section presents the research methodology employed for autonomous CAN reverse engineering in this paper. The process begins with attaching a system to the vehicle and recording data while it is operated. Simultaneously, the system captures the vehicle's movement and records all of the CAN data generated during the drive. The collected data is then processed, and the proposed techniques are utilized to identify the specific CAN channels associated with the vehicle's accelerator and brake pedals. The key components and procedures of this methodology are discussed in detail below, including interfacing with the vehicle's CAN bus, measuring and interpreting vehicle actions using an IMU sensor, parallel sampling using the Robotic Operating System (ROS), CAN channel organization, correlating CAN channels to vehicle actions using the Pearson correlation coefficient, and identifying CAN channels relating to vehicle controls. Additional recordings taken while the vehicle is stationary are used to further aid in identifying the complete range of the controls.

*3.1. Method for Autonomous CAN Reverse Engineering*

The method used for autonomous CAN reverse engineering is discussed in this section. It starts by explaining the process of interfacing with the vehicle CAN bus through the OBD-II port using a commercial CAN bus analyzer tool. The IMU sensor, which helps measure vehicle actions by recording 3D digital accelerations and angular velocities, is introduced. Parallel sampling is achieved through the use of the Robotic Operating System (ROS) as middleware, enabling modular software development and data recording. The organization of CAN channels is discussed, as are different configurations and tokenizations. Finally, the correlation of CAN channels to vehicle actions is examined using the Pearson correlation coefficient algorithm. The identification of CAN channels related to vehicle controls is discussed, including the aid provided by additional recordings taken while the vehicle is stationary.

3.1.1. Interfacing with Vehicle CAN Bus

The raw CAN bus data (see [30], [32], [33]) is accessed via the vehicle's on-board diagnostic (OBD-II) port. This port is often located below the vehicle's steering wheel. To interface with the raw data feed, a commercial off the shelf CAN bus analyzer tool [35] is used. This tool translates raw CAN data into serial data over USB. The drivers and software associated with this device are described in Section 3.2.4. Additionally, an OBD-II to DB9 cable is used to connect the CAN bus analyzer tool to the vehicle OBD-II connector.

3.1.2. Measuring and Interpreting Vehicle Actions

An IMU sensor [36], including an accelerometer to measure 3D digital accelerations and a gyroscope to measure 3D angular velocities, is used. It identifies the sudden movements of the vehicle caused by the accelerator and brake pedals. To achieve this, the system records all IMU data for a set duration, which is then analyzed to determine the vehicle's behavior during this timeframe. This movement data is correlated with the relevant CAN data to help identify the CAN channels associated with vehicle actions.

Vehicle Acceleration - Under normal driving conditions, when the operator engages the accelerator pedal, the vehicle accelerates at a specified rate. This rate of acceleration is measurable by the accelerometer and is directly correlated with the magnitude of depression of the accelerator pedal [3]. If the operator presses the pedal down further, the vehicle accelerates faster. There is a direct relationship between the actuation of this pedal and the IMU sensor data, which makes it a suitable candidate for time series correlation analysis.

Vehicle Deceleration - Deceleration is expected when the vehicle is in motion and the operator engages the brake pedal. Like with acceleration, this action has a linear relationship with the data collected by the IMU sensor. When the vehicle is in motion and the operator presses the brake pedal, the vehicle is expected to slow down. Like with vehicle acceleration, there is a linear relationship between how far the brake pedal is depressed and the magnitude of deceleration observed by the IMU sensor.

3.1.3. Parallel Sampling

The Robotic Operating System (ROS) was used as middleware for this system. ROS uses a data-centric architecture, allowing the software used for the research to be written in a modular fashion which further enhances its scalability. The IMU system and CAN system each have their own respective ROS nodes to communicate with the ROS system. These ROS nodes are executed in their own threads on the computer that is used to control and monitor this system. ROS includes native features and tools that allow data within the ROS network to be recorded over a shared time series. These recordings are serialized for post processing.

3.1.4. CAN Channel Organization

In Section 2.2, the concept of a CAN channel was introduced. A CAN channel includes the arbitration ID of a CAN frame and a specific portion of the CAN frame's payload. Reverse engineering and identifying the portions of the CAN frame payload relating to a CAN channel has already been researched [5], [10], [20]. It is important note that processing time is not a constraint that is applied in this research, as reverse engineering is not being conducted in real time. Therefore, an exhaustive exploration approach can be used. The CAN frame can be broken down into many different possible CAN channels. Figure 5 and 6 show how CAN frames can be divided into individual CAN channels. Each CAN channel's data is placed at a designated start location within the CAN payload. The baseline configuration uses a byte format (eight bits) length with consecutive start and end bits. The next configuration uses nine-bit lengths. Some implementations use least-significant bit (LSB) format whereas others use most-significant bit (MSB) format. In both cases, the software expands the length of each channel (leftwards in the case of MSB and rightwards in the case of LSB) until the maximum length of sixteen bits is reached. This creates many possible combinations of different CAN channel tokenizations.

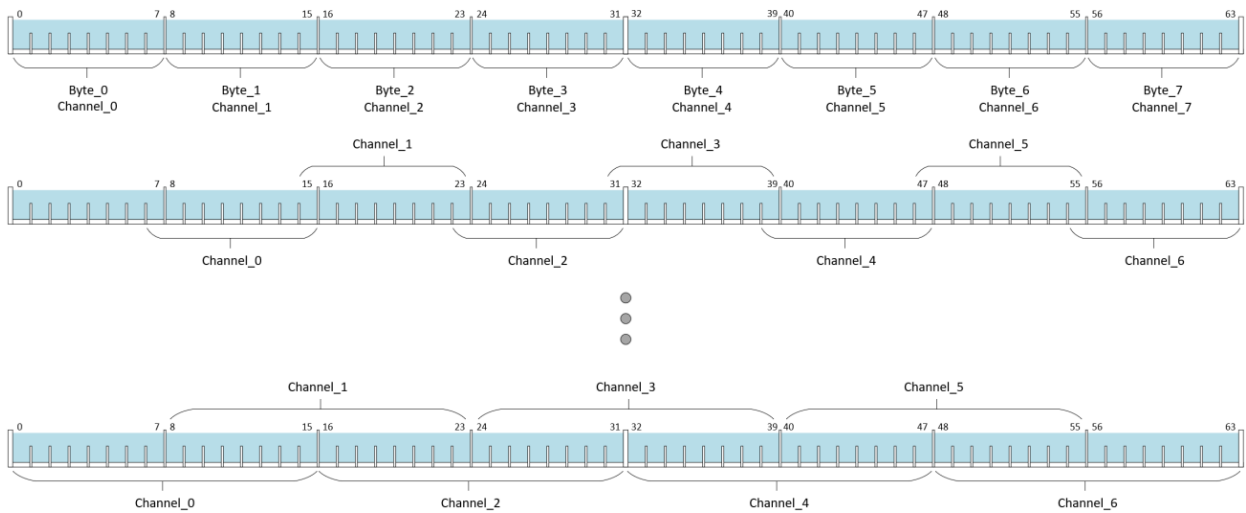

**Figure 5.** CAN channel (MSB) configurations.

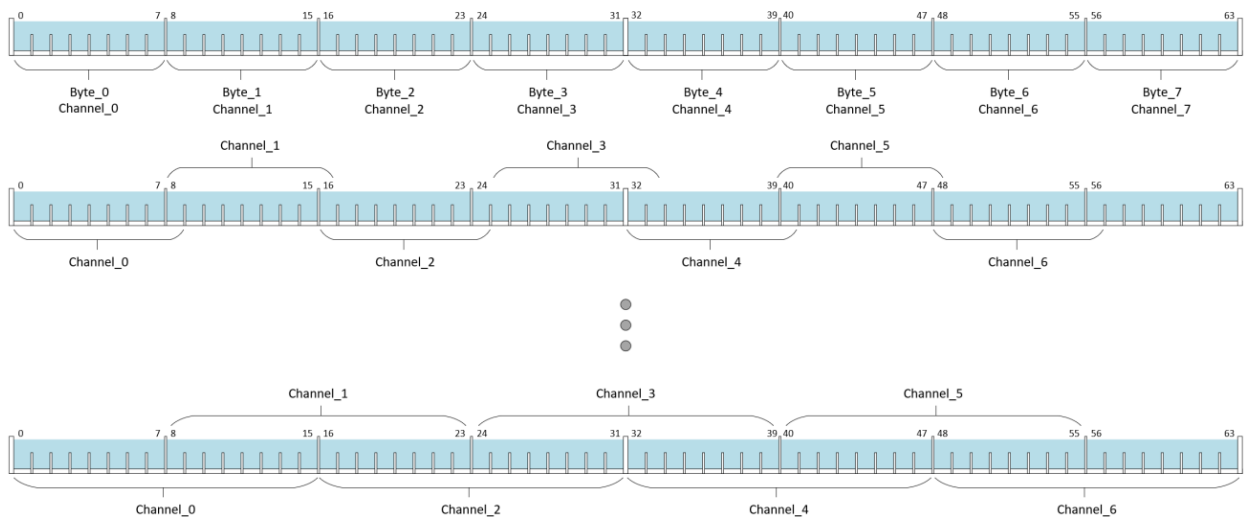

**Figure 6.** CAN channel (LSB) configurations.

3.1.5. Correlating CAN Channels to Vehicle Actions

The Pearson correlation coefficient algorithm is a common method for determining the correlation between signals (data) over a shared time series. It is a suitable solution for measuring the relationship between inertial/displacement data and CAN Channel data over time. It was previously used by Blaauwendraad and Kieberl [22] for correlating between known OBD-II PIDs and raw CAN data. Using the algorithms described subsequently, this process produces two separate datasets, including the CAN channels that relate to vehicle acceleration and declaration. The equation used for Pearson correlation is [22]:

$$r_{xy} = \frac{\sum_{i=1}^{n}(x_i - \bar{x})(y_i - \bar{y})}{\sqrt{\sum_{i=1}^{n}(x_i - \bar{x})^2}\sqrt{\sum_{i=1}^{n}(y_i - \bar{y})^2}} \tag{1}$$

Blaauwendraad and Kieberl used the algorithm with *x* being the list of values returned from OBD-II requests, and *y being the list of average values for a specified CAN channel*. In this research, a modification is made where *x* is a series of IMU related information and *y is the raw data from a specified CAN channel*. The specified CAN channel is determined using the method discussed in the previous section. This is used in the rate of change correlation algorithm which is discussed later.

3.1.6. Identifying CAN Channels Relating to Vehicle Controls

Once CAN message correlation to vehicle actions is complete, the resulting datasets are further analyzed to identify which CAN channels pertain to the accelerator and brake pedals. A detailed explanation of the process of the vehicle controls discovery algorithm is provided later.

To aid this process, additional recordings are taken while the vehicle is not in motion. These recordings are intended to capture a range of CAN messages while the driver engages the accelerator and brake pedals separately. These recordings help ensure that the full range of the controls are measured, as typically the controls are not fully engaged (to reach their maximum values) during vehicle operations.

*3.2. System Hardware and Drivers*

In this section, the hardware components and software used to provide the system's functionality are described. The section begins by discussing the selection and features of the inertial measurement unit (IMU) sensor, including its integrated capabilities such as USB communication, ROS system source code, and built-in algorithms. It then explores the IMU drivers and algorithms employed, along with the development of a dedicated ROS node for IMU sampling and publishing. Then it delves into the CAN interface module utilized to connect with the vehicle's CAN bus. This is followed by an explanation of the CAN software, drivers, and ROS integration. Finally, the section presents details about the computer and operator interface, including the hardware specifications and software tools used for recording and analysis. Figure 7 displays the equipment used in this work.

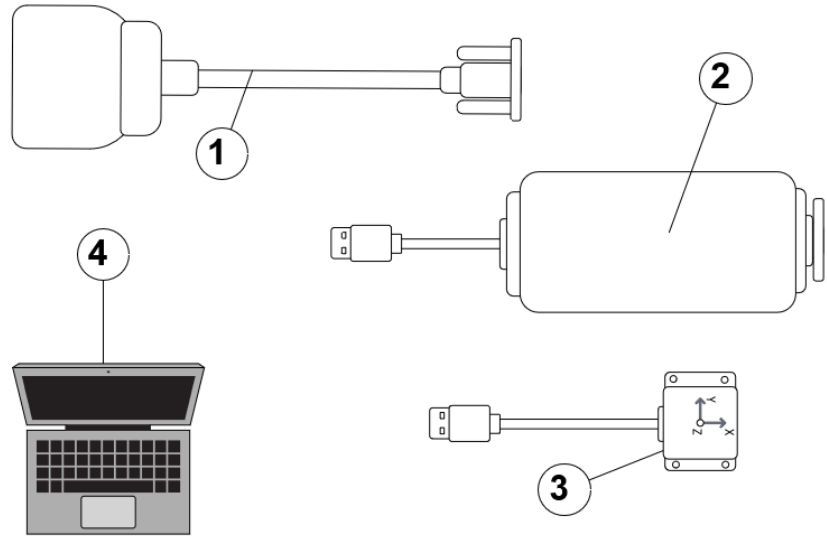

| Item | DESCRIPTION | Part Number / Brand |
|------|-------------|---------------------|
| 1 | OBD-II – to – DB9 Harness | 1DB9-OBD-A / Innomaker |
| 2 | CAN-USB Analyzer Tool | APGDT002 / Microchip |
| 3 | IMU Sensor (10 dof) – to USB | CMP10A - Yahboom |
| 4 | Laptop | Precision 5530 / Dell |

**Figure 7.** System equipment.

3.2.1. Inertial Measurement Unit

The IMU sensor [36] was chosen due to its features such as having a USB communication, providing ROS system source code, and including a built-in Kalman filter algorithm that can output Euler angles (roll, pitch, yaw), quaternion, position, velocity, acceleration, angular velocity, and a magnetic field vector.

IMU drivers and algorithms - Open-source drivers and example software, including ROS integration, for the COTS IMU module [37] were used. The IMU sensor, by default, communicated at 10Hz; however, the company also provides software [37] to alter the IMU sensor's internal sample rate and publishing frequencies. For this application, a 100Hz sample rate was used.

IMU node (ROS integration) - A ROS node was developed to handle IMU sampling and publishing to the local ROS network. The IMU ROS node listens for new updates from the IMU module and, anytime a new value is presented, the IMU ROS node publishes this information to the ROS network.

3.2.2. IMU Recording and Data Organization

Software was developed to record the IMU data as it is transmitted onto the ROS network. To access the published IMU data, this software subscribes to the IMU ROS topic "IMU_Filtered". It then creates a comma separated value (CSV) file to record the IMU data into. Every time a new data sample is available on the ROS network, the IMU recording software takes the sample and adds the information to an appropriate column. An example of this format is shown in the table below. Upon conclusion of the recording, the IMU recording software saves the CSV file to the location selected by the operator. The IMU CSV file can now be used for correlation analysis.

**Table 3.** IMU Data Organization.

| Time Stamp (seconds) | Linear Y |
|---:|---:|
| 0 | 0.886614 |
| 0.0200396 | 0.974758 |
| 0.040128 | 0.975089 |
| 0.0600224 | 0.89549 |
| 0.0799887 | 0.874468 |

3.2.3. Controller Area Network

A CAN interface module was used to interface with the vehicle's CAN bus via the OBD-II connector. This interface was used to record CAN data during the specific acceleration events. The CAN interface chosen for this system was the ADGDT002 CAN bus analyzer tool provided by Microchip [35]. On one end of the analyzer tool is a universal serial bus (USB) connector, intended to be connected to a computer, and the other end has a DB9 connector to interface with the vehicle's high-speed CAN system.

3.2.4. CAN Software Description

This section describes the software which was used to collect and analyze CAN data. Two different software components were used.

CAN analyzer driver software - To interconnect with the ADGDT002 hardware, the laptop needed an additional Linux kernel driver [38]. Originally the Microchip CAN analyzer was only supported in the Windows environment; however, the open-source SocketCAN driver [39] allows the tool to be used in Linux. SocketCAN integrates into the networking stack on the device.

CAN node (ROS integration) software - A package, within the ROS library, called "socketcan_bridge" [40] was used for interconnecting ROS with the SocketCAN network stack. The socketcan_bridge package includes a ROS node called socketcan_bridge_node, that was used to translate messages between the CAN environment and the ROS environment. With the socketcan_bridge_node, CAN frames received on the SocketCAN device (ADGDT002) are published to a ROS topic called "*received_messages*" and ROS messages received on the "*sent_messages*" ROS topic are sent to the SocketCAN device (ADGDT002). Once the SocketCAN network stack has been instantiated on the device, the socketcan_bridge_node can be run within the ROS environment to connect to the ADGDT002 connected to the vehicle's CAN bus, thus interconnecting the CAN interface to the ROS network.

3.2.5. CAN Recording and Data Organization

The ROS CAN node publishes all CAN messages received by the CAN module. CAN recording software was developed to organize the CAN messages into their own sections sorted by their CAN frame ID. Typically, all associated data is included in the payload of one message for a given arbitration ID. Automotive CAN uses a data length code (DLC) of up to eight bytes. Initial testing showed that there were multiple CAN frames that had a DLC of less than eight bytes. To handle this, each CAN frame records eight bytes of information. The bytes that were not included within the CAN frame's DLC are set to zero, and an additional column was created to store the DLC of the CAN frame.

To access the published CAN data, the CAN recording software was subscribed to the CAN ROS topic "received_messages". Since each CAN frame has its own unique time series data, a CSV file was created for every new arbitration ID and all subsequent frames with this ID are placed in this file. The CAN recording software is designed to move back and forth between CSV files as new CAN messages are received. Upon conclusion of the recording, the CAN recording software closes and saves the CSV files in the operator-selected directory. An example of the CAN data is shown in Table 4. The data within the bytes are represented as integer values.

**Table 4.** CAN data organization (CAN frame ID: 272).

| Arbitration ID: 272 | | | | | | | | |
|---|---|---|---|---|---|---|---|---|
| Time Stamp (seconds) | Byte 0 | Byte 1 | Byte 2 | Byte 3 | Byte 4 | Byte 5 | Byte 6 | Byte 7 |
| 0.00338745 | 0 | 10 | 173 | 0 | 0 | 19 | 0 | 80 |
| 0.0158813 | 0 | 10 | 129 | 0 | 0 | 19 | 0 | 56 |
| 0.0288999 | 0 | 10 | 129 | 0 | 0 | 19 | 0 | 0 |
| 0.0410981 | 0 | 10 | 168 | 0 | 0 | 19 | 0 | 104 |
| 0.0533037 | 0 | 10 | 168 | 0 | 0 | 19 | 0 | 80 |
| … | … | … | … | … | … | … | … | … |

3.2.6. Computing System

A Dell Precision 5530 laptop running the Ubuntu 20.01 Linux operating system was used to run the system described herein. This laptop has an Intel Core i9 8$^{th}$ generation processor and 32GB of RAM. It hosted all of the associated software nodes for making measurements, processing data, performing reverse engineering, and presenting the information to the operator.

Operator Interface Software - Linux shell terminals are utilized to interface with the software used for the project. To allow visibility of all of the concurrently running software during recording and analysis, the Linux application Terminator [41], [42] was used to provide multiple shell windows (in separate threads). Within Terminator, the operator can open a separate shell window for each of the applications and scripts within the project. During research and development activities, ROS is interacted with through a shell terminal, and specific commands. To simplify the initialization and execution of the system, these shell commands were stored in a Linux Bash file. The operator executes this bash file from within a shell terminal, and the ROS nodes and initialization occur as programmed.

Multiple ROS nodes were launched using a feature called ROS Launch [43]. A launch file initializes multiple ROS nodes and can assign specific parameters to each node, if desired. This feature is a convenient way to configure the system and start multiple ROS node applications simultaneously. Several ROS Launch calls are included in a single Linux Bash file for this system.

Creating recordings of the ROS environment - To make recordings, the user opens a new terminal and runs specific Bash commands to set up the shell as a "recording node". Using the appropriate ROS commands, the user creates a ROSbag that subscribes to the desired ROS topics for recording. To do this, a Bash file that contains the necessary ROS commands is called with an argument that specifies the directory where the recorded data will be saved. This provides a recording of all of the information published to the ROS network by topic. This file can be used for playback to enable research activities.

3.3. Recording Data for Reverse Engineering

The software algorithms to reverse engineer CAN data require two different types of recordings. For each of these recordings, the recording system is configured as shown in Table 5, using the hardware shown in Figure 7. The recording software, ROS, and drivers associated with the IMU and CAN peripherals were discussed in previous sections. Software setup and initialization was automated using software scripts that are run in a terminal application [41], [42] on the laptop, as previously discussed.

**Table 5.** Recording Configuration.

| Number | Configuration Verification |
|---|---|
| 1 | OBD-II – to – DB9 cable is securely connected to the OBD-II interface within the vehicle. |
| 2 | The CAN-USB analyzer tool is securely connected to the DB9 interface on the OBD-II cable and is also connected to the test laptop via USB. |
| 3 | IMU sensor is securely positioned within the vehicle in such a way that it will not move during vehicle operation and is connected to the test laptop via USB. |
| 4 | The test laptop is powered on, with the software prepared for recording IMU and CAN data. |

The two types of recordings that are required are the trip and calibration recordings. Each is now discussed.

3.3.1. Trip Recording

For the process of trip recording, the operator drives for a short period of time, while the system logs all IMU and CAN information. Once the trip is completed, the operator terminates the software application, and the datasets are analyzed using the algorithms discussed in the subsequent sections. It was found that a 5 to 10-minute drive, for each vehicle, was sufficient, as it allowed enough vehicle acceleration and deceleration events to be recorded. When taking a recording, the system was configured as discussed in Table 5 and was running the recording software for the IMU and CAN systems.

3.3.2. Calibration Recording

Calibration recordings are collected to aid in identifying the CAN channels associated with the accelerator and brake pedals. This methodology is similar to LibreCAN's approach [20]; however, there are some distinctions. LibreCAN employed a special recording method to detect body-related CAN information related to door, window and light operations and other activities. In contrast, this work employs a recording approach where the operator depresses the accelerator and brake pedals to their fullest extent without the engine running.

A separate recording for each control input is collected. The vehicle is parked and, for maximum isolation of the control inputs, the vehicle's engine is powered off; however, it is important that the vehicle's electronics remain powered on and communicating. Vehicles with a keyless (push button) start, were found to send CAN messages if the driver had the remote key with them while sitting in the driver's seat. The vehicle was turned on and then turned off again, using the push button interface. It was found that, if the driver did not open the door, the vehicles internal ECUs would still communicate over CAN. For vehicles that had a traditional keyed starting system, the key was turned to the ignition stage (not start), allowing the electronics to initialize and communicate without the engine running.

3.3.2.1. Accelerator Pedal Recording

The accelerator pedal recording involved pushing the accelerator pedal down multiple times to gather all associated CAN data, without the engine running or the vehicle moving. To get the best results, the accelerator pedal needs to actuate to its extremes and be pushed all the way down and released all the way up. With the vehicle in park, and the engine in the off condition, the driver began the accelerator pedal recording by initiating the recording system and actuating the pedal a few times. Once complete, the operator stopped the software application and the process is complete.

3.3.2.2. Brake Pedal Recording

Much like the accelerator pedal recording, the goal of the brake pedal recording was to record all of the CAN traffic while actuating the brake pedal. The configuration of the vehicle and the recording system was identical to the accelerator pedal except for labeling the recording as for the brake pedal. With the vehicle in park, and the engine in the off condition, the operator began the brake pedal recording by initiating the recording system and actuating the pedal a few times. Since the vehicle is not running, the brake lines will likely build up pressure, so it will become difficult to push the brake down further after the first actuation. Once several actuations were performed, the operator ended the software application and the process was complete.

*3.4. System Software Overview*

Two main algorithms were developed to perform the automotive CAN reverse engineering presented herein. The first algorithm identifies unique correlations between IMU data and organized CAN channels. The second algorithm takes the CAN channels' output from the first algorithm and identifies which of those CAN channels specifically relate to each vehicle control. Figure 8 provides an overview of the software used for this work. This software is discussed further in the following subsections.

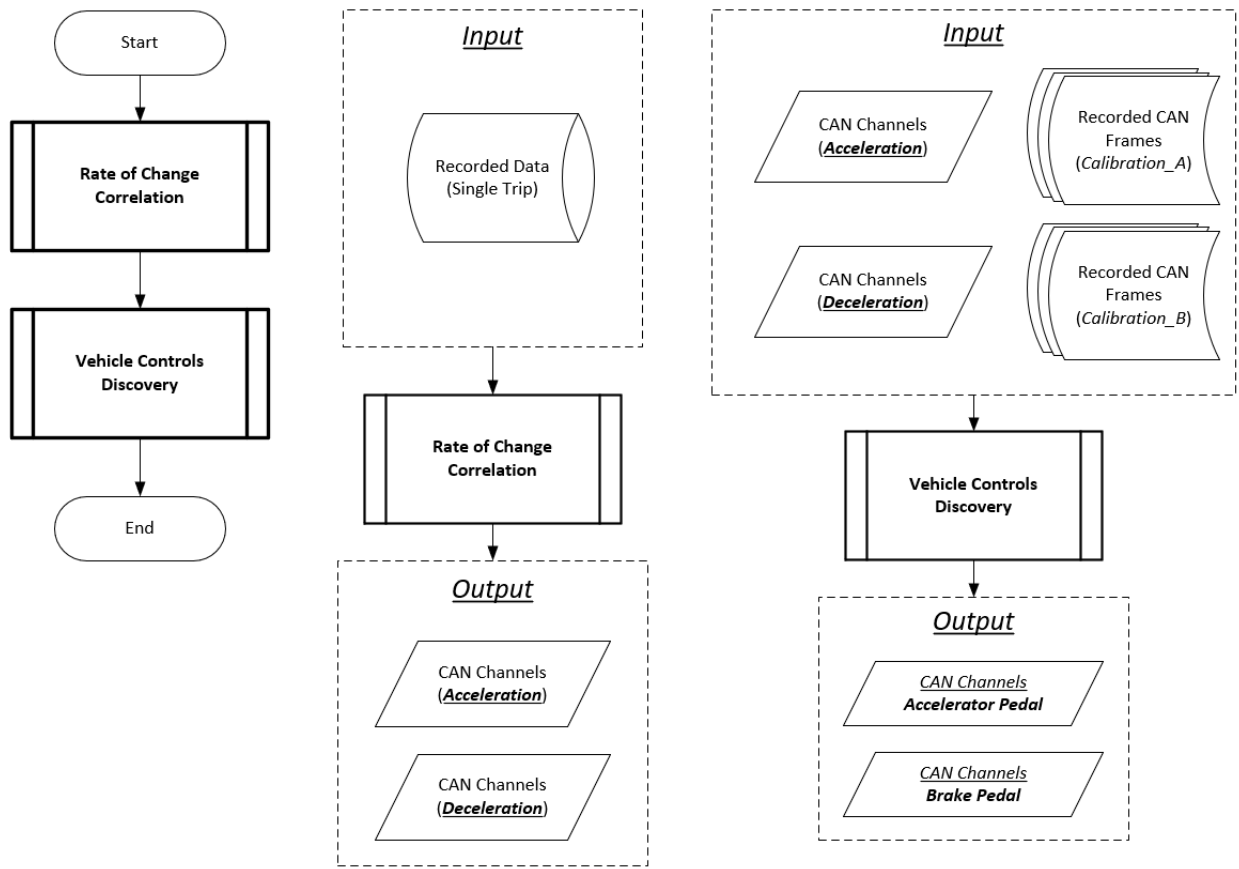

**Figure 8.** System software overview.

*3.5. Rate of Change Correlation Software*

Once a recording has been collected, the rate of change correlation algorithm is used to analyze the data and reverse engineer the CAN channels relating to acceleration and deceleration. This algorithm analyzes the entire time series recording, identifying correlations between the recorded IMU data and the recorded CAN channels. The output of this algorithm is a list of identified CAN channels that correlate most strongly to vehicle acceleration and deceleration. The results from running this algorithm are presented in Section 5 and analyzed in Section 6.

Figure 9 presents a flowchart for the rate of change correlation algorithm that was developed for this work.

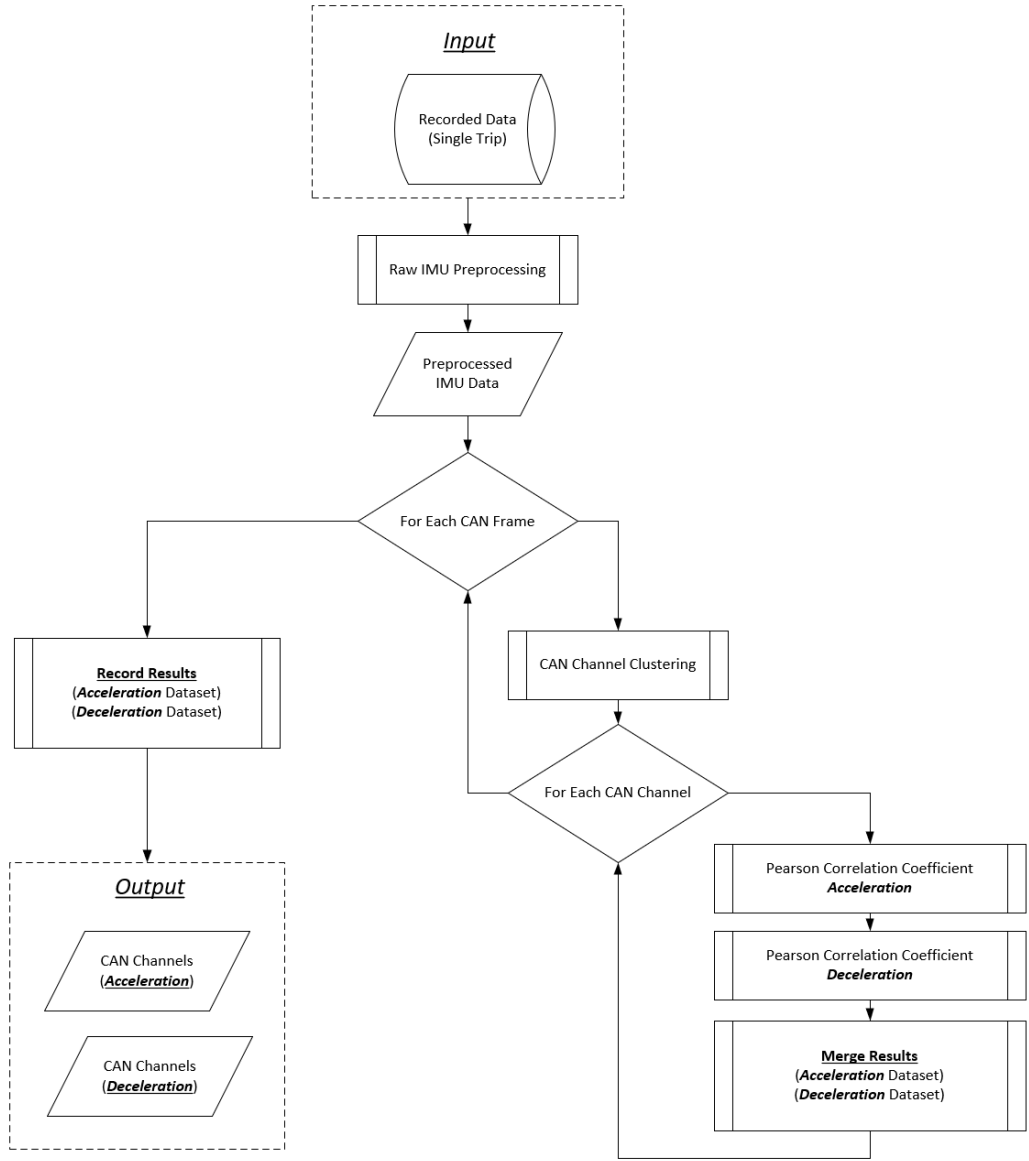

**Figure 9.** Rate of change correlation algorithm flow chart.

The algorithm starts by collecting and organizing the IMU data during the raw IMU preprocessing phase. During this phase, the IMU directional data is categorized as acceleration and deceleration measurements, to later merge with CAN channel data for correlation.

Next, a nested FOR loop iterates through all of the recorded CAN messages. The first FOR loop separates the recorded CAN messages into a set of individual CAN frame series, where each arbitration ID has its own series. For each CAN frame series, the CAN channel clustering process breaks down each CAN frame into individual CAN channels. This creates many possible combinations, providing coverage of different CAN channel tokenizations, which are then processed in an exhaustive manner.

The second FOR loop merges the CAN channels with the IMU acceleration and deceleration datasets and performs the Pearson Correlation algorithm. The Python Pandas [45] library's "merge_asof" function was used to merge the two data frames. The smaller data frame is the dominant one in the merger with data being associated with the closest time match in the other set. The Pearson Correlation Coefficient is calculated for each resulting dataset and provides correlation coefficients between vehicle acceleration and deceleration and each CAN channel.

Finally, the algorithm appends the CAN channel (acceleration and deceleration) information for each resulting dataset. When the FOR loop completes iterating through all of the CAN frames, the results are recorded and provided as an input to the vehicle controls discovery algorithm which is described in the next section.

*3.6. Vehicle Controls Discovery Algorithm*

The vehicle controls discovery algorithm is run on both datasets returned from the rate of change correlation algorithm. For each dataset (CAN channels – acceleration and CAN channels – deceleration), the algorithm searches and identifies the CAN channels that relate to the controls for that type of vehicle movement. The algorithm achieves this by processing each of the strongly correlated CAN channels and analyzing the specific CAN channel data from the calibration associated with the dataset (i.e., acceleration = accelerator pedal and deceleration = brake pedal). The algorithm converges on a list of CAN channels that are most likely relate to the respective vehicle control.

Figure 10 presents a flowchart for the vehicle controls discovery algorithm that was developed for this work.

The algorithm starts with a FOR loop that analyzes each CAN channel returned from the rate of change correlation algorithm. Inside the loop, the algorithm processes the time series data for the specified CAN channel during its calibration recording for either the accelerator or brake pedal, as relevant. It then calculates the number of unique values and the range of values for the series of data. The derivative of the time series and the standard deviation of the derivate series are computed. This helps to identify how the CAN channel's data changes over time, as a pedal sensor is expected to be smooth and consistent in its increments. The algorithm quantifies this by calculating the quotient between the standard deviation of the derivative series and the range of values the CAN channel includes.

Once all input CAN channels have been analyzed, a final sorting process takes place. The CAN channels with the least number of unique values are removed, and the remaining channels are sorted based on how smoothly they changed over time. Only a subset of CAN channels are returned. The number of channels can be adjusted to fit the specific needs of the user.

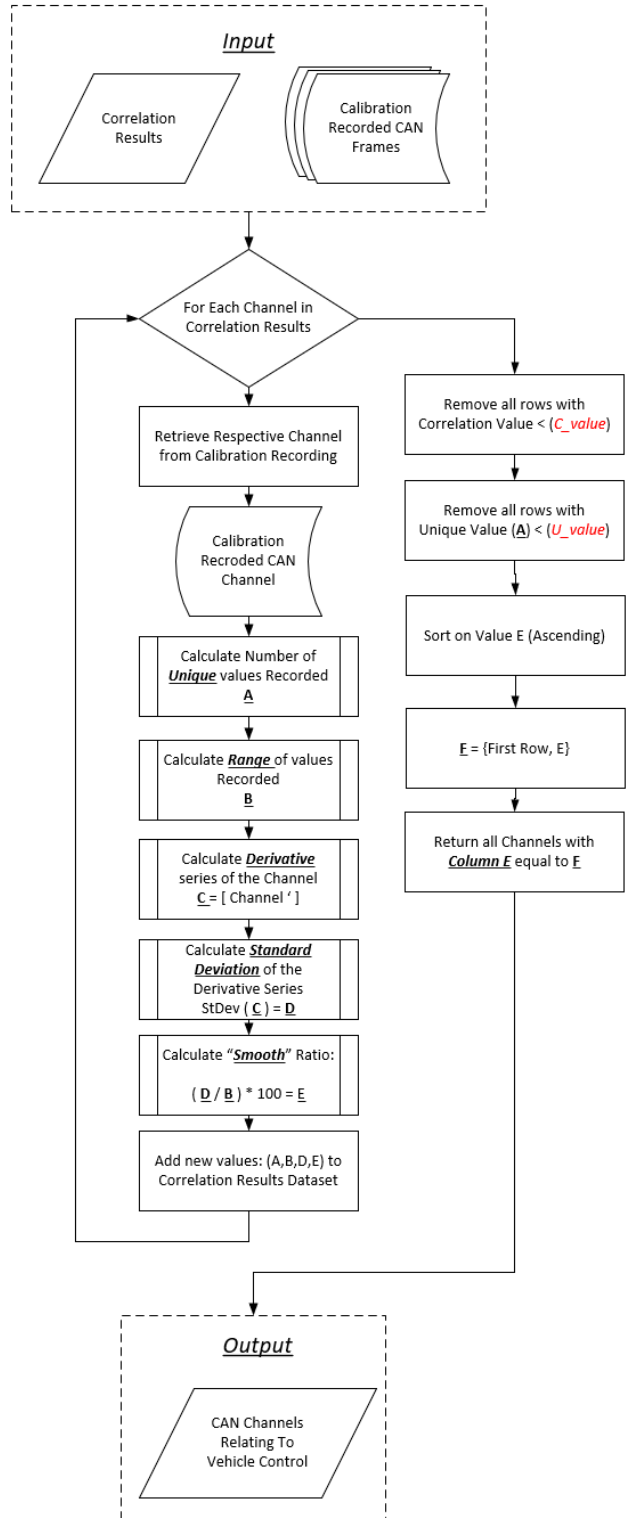

**Figure 10.** Vehicle controls discovery algorithm flowchart.

## 4. Results and analysis

In this section, the data collected and results generated are presented. The subsections of this section discuss the reverse engineering process and the outputs of the various algorithms using recordings taken with the main test vehicle, a 2016 GMC Sierra. To assess the efficacy of the reverse engineering system, the results from multiple recordings with this same vehicle are presented. Multiple other vehicles were tested to demonstrate the broad utility of the system and its scalability. The results from these vehicles are summarized in Appendix A.

Throughout this section, references to specific CAN channels are often made. The nomenclature used to present this is straight forward. For example, "211_msb_sixteen_bit_0" means an ID of 211, a bit length of 16, a channel of 0 and an endian style of MSB. The generic format is: (*Frame ID)_(endianness)_(channel bit length)_bit_(channel number in frame)*. However, if the bit length is eight bits, the method's default channel size, the format is simplified to: *(frame_ID)_byte_(byte number)*.

*4.1. Acceleration Reverse Engineering*

In this section, the reverse engineering results for the accelerator pedal are presented. The IMU sensor was positioned so that the accelerations in the Y axis corresponded to the vehicle's forward and aft movements. Negative values measured by the IMU's Y axis represent forward vehicle acceleration. As explained above, the Y axis of the IMU sensor underwent filtering to separate the vehicle's acceleration signals as a distinct signal. Figure 11 displays the data recorded during a trip with the primary test vehicle, a 2016 GMC Sierra.

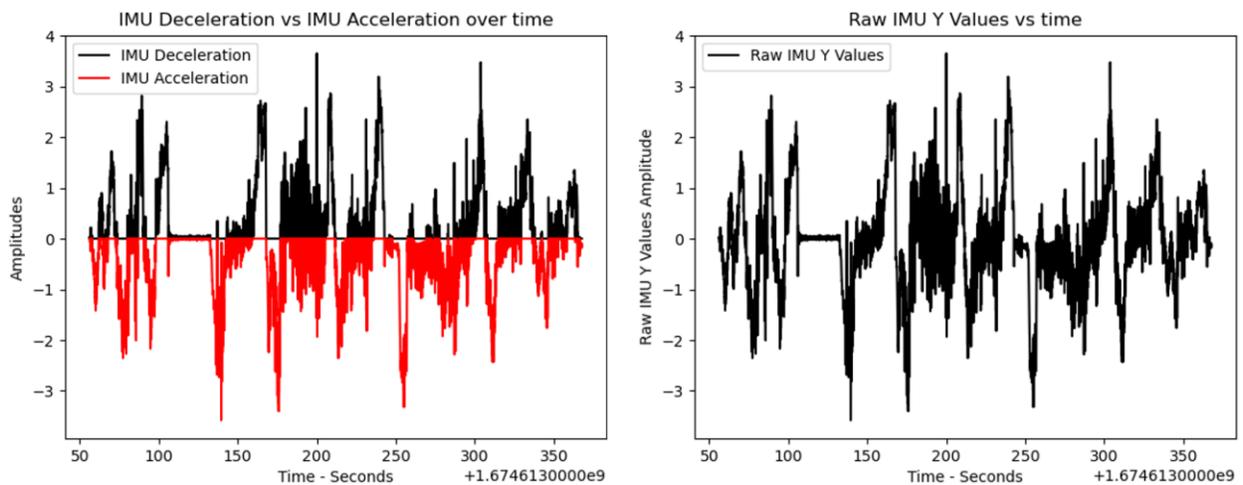

**Figure 11.** IMU acceleration values.

The rate of change correlation algorithm used the acceleration signal as a baseline for correlation with all of the CAN channels recorded during the drive. For every CAN channel, a percent correlation was calculated to determine how closely it correlates with vehicle acceleration. The results from this correlation that related most strongly to vehicle acceleration are shown in Table 6.

**Table 6.** Acceleration results.

| ID | Channel | Correlation |
|---|---|---|
| 211 | msb_fifteen_bit_0 | 0.87761497 |
| 211 | msb_fourteen_bit_0 | 0.87761497 |
| 211 | msb_sixteen_bit_0 | 0.87761497 |
| 211 | msb_thirteen_bit_0 | 0.87761497 |
| 170 | msb_fifteen_bit_0 | 0.87734619 |

| 170 | msb_thirteen_bit_0 | 0.87734619 |
| 170 | msb_fourteen_bit_0 | 0.87734619 |
| 170 | msb_sixteen_bit_0 | 0.87734619 |
| 453 | msb_sixteen_bit_0 | 0.86757076 |
| 453 | msb_fifteen_bit_0 | 0.86757076 |
| 453 | msb_thirteen_bit_0 | 0.86757076 |
| 453 | msb_fourteen_bit_0 | 0.86757076 |
| 170 | byte_0 | 0.86354249 |
| 211 | byte_0 | 0.86336653 |
| 454 | msb_fifteen_bit_6 | 0.85032099 |
| 454 | msb_sixteen_bit_6 | 0.85032099 |
| 454 | msb_sixteen_bit_2 | 0.85030932 |
| 454 | msb_fifteen_bit_2 | 0.85030932 |
| 454 | msb_fourteen_bit_2 | 0.85030932 |
| 454 | msb_thirteen_bit_2 | 0.85030932 |
| 453 | byte_0 | 0.84890293 |
| 454 | msb_thirteen_bit_5 | 0.84839341 |
| 454 | msb_eleven_bit_5 | 0.84839341 |
| 454 | msb_ten_bit_5 | 0.84839341 |
| 454 | msb_nine_bit_5 | 0.84839341 |

The channels associated with CAN IDs 211 and 170 demonstrated the strongest correlation with acceleration. The channel associated with ID 211 was 211_msb_sixteen_bit_0, and the channel associated with ID 170 was 170_msb_sixteen_bit_0. The graphs in Figure 12 and Figure 13 display the values of each channel over the course of a drive, in relation to acceleration.

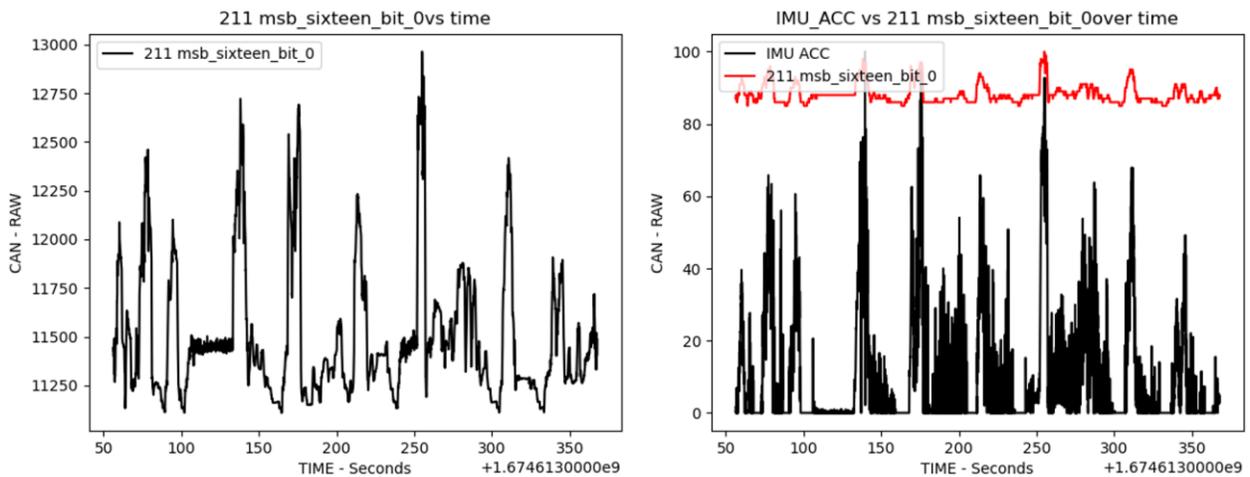

**Figure 12.** Acceleration reverse engineering CAN channel 211_msb_sixteen_bit_0.

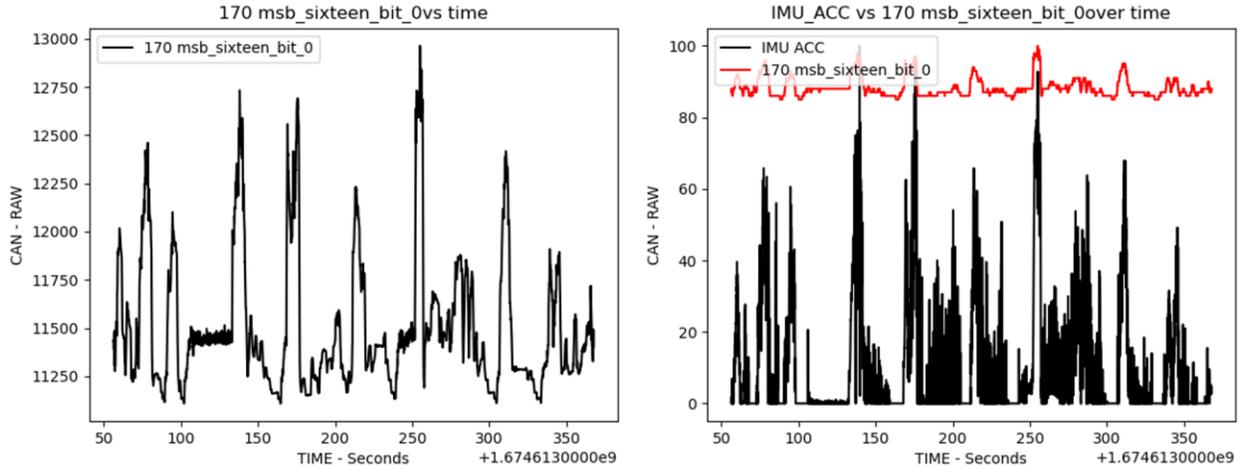
**Figure 13.** Acceleration reverse engineering CAN channel 170_msb_sixteen_bit_0.

*4.2. Accelerator Pedal Reverse Engineering*

The time series data from the accelerator pedal calibration recording was graphed. This is shown in Figure 14. IDs 211 and 170 both had strong correlations with acceleration; however, their values remained constant during the recording of the accelerator pedal calibration. This shows the necessity of the vehicle controls discovery algorithm to identify which of the acceleration CAN channels relate to the accelerator pedal. Using this, these two highest-correlating IDs can be discounted as not representing the accelerator pedal.

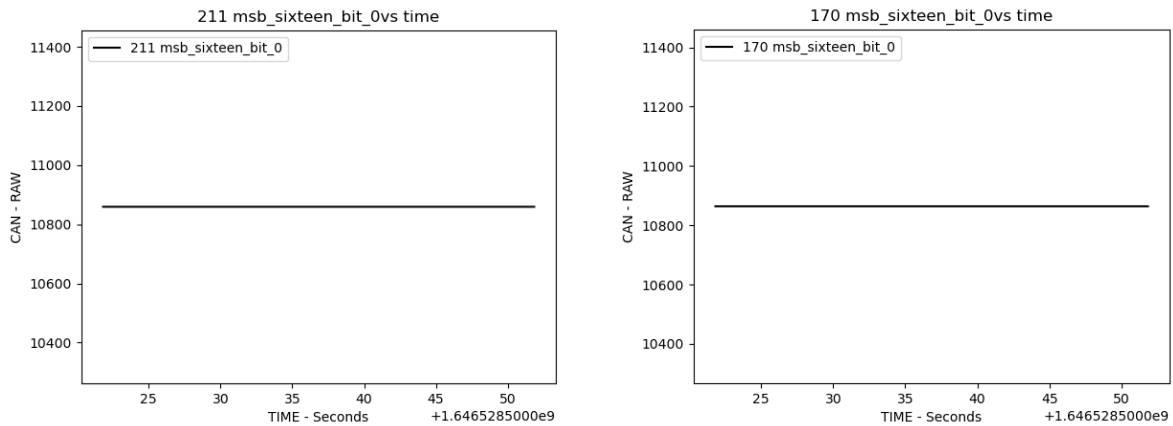
**Figure 14.** Accelerator pedal calibration verification.

Figure 15 illustrates the anticipated waveform that should be generated for the accelerator pedal being depressed and released during its calibration recording.

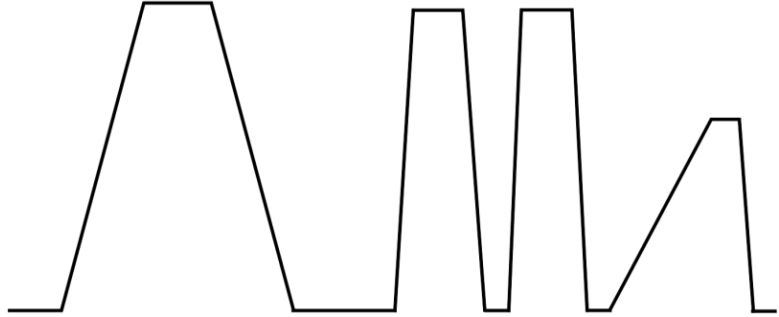

**Figure 15.** Anticipated accelerator pedal waveform.

The vehicle controls discovery algorithm takes the dataset from the acceleration correlation results and adds additional calculated values for each CAN channel. The algorithm processes the accelerator pedal recording and identifies the range of values (shown in the range column in Table 7) and number of unique values (shown in the unique column) for each of the CAN channels in the acceleration correlation results. Table 7 shows the top 25 channels of this data.

**Table 7.** Appending range and unique columns for acceleration channels.

| ID | Channel | Correlation | Range | Unique |
|---|---|---|---|---|
| 211 | msb_fifteen_bit_0 | 0.87761497 | 0 | 1 |
| 211 | msb_fourteen_bit_0 | 0.87761497 | 0 | 1 |
| 211 | msb_sixteen_bit_0 | 0.87761497 | 0 | 1 |
| 211 | msb_thirteen_bit_0 | 0.87761497 | 0 | 1 |
| 170 | msb_fifteen_bit_0 | 0.87734619 | 0 | 1 |
| 170 | msb_thirteen_bit_0 | 0.87734619 | 0 | 1 |
| 170 | msb_fourteen_bit_0 | 0.87734619 | 0 | 1 |
| 170 | msb_sixteen_bit_0 | 0.87734619 | 0 | 1 |
| 453 | msb_sixteen_bit_0 | 0.86757076 | 4521 | 126 |
| 453 | msb_fifteen_bit_0 | 0.86757076 | 4521 | 126 |
| 453 | msb_thirteen_bit_0 | 0.86757076 | 7979 | 126 |
| 453 | msb_fourteen_bit_0 | 0.86757076 | 16171 | 126 |
| 170 | byte_0 | 0.86354249 | 0 | 1 |
| 211 | byte_0 | 0.86336653 | 0 | 1 |
| 454 | msb_fifteen_bit_6 | 0.85032099 | 32262 | 108 |
| 454 | msb_sixteen_bit_6 | 0.85032099 | 65030 | 108 |
| 454 | msb_sixteen_bit_2 | 0.85030932 | 4093 | 77 |
| 454 | msb_fifteen_bit_2 | 0.85030932 | 4093 | 77 |
| 454 | msb_fourteen_bit_2 | 0.85030932 | 16257 | 77 |
| 454 | msb_thirteen_bit_2 | 0.85030932 | 8065 | 77 |
| 453 | byte_0 | 0.84890293 | 18 | 19 |
| 454 | msb_thirteen_bit_5 | 0.84839341 | 64 | 44 |
| 454 | msb_eleven_bit_5 | 0.84839341 | 64 | 44 |

| ID | Channel | Correlation | Range | Unique |
|---|---|---|---|---|
| 454 | msb_ten_bit_5 | 0.84839341 | 64 | 44 |
| 454 | msb_nine_bit_5 | 0.84839341 | 64 | 44 |

The vehicle controls discovery algorithm then removes the CAN channels that appear constant during the accelerator pedal recording. The top 25 results after this are shown in Table 8.

**Table 8.** Reducing acceleration CAN channels.

| ID | Channel | Correlation | Range | Unique |
|---|---|---|---|---|
| 453 | msb_sixteen_bit_0 | 0.86757076 | 4521 | 126 |
| 453 | msb_fifteen_bit_0 | 0.86757076 | 4521 | 126 |
| 453 | msb_thirteen_bit_0 | 0.86757076 | 7979 | 126 |
| 453 | msb_fourteen_bit_0 | 0.86757076 | 16171 | 126 |
| 454 | msb_fifteen_bit_6 | 0.85032099 | 32262 | 108 |
| 454 | msb_sixteen_bit_6 | 0.85032099 | 65030 | 108 |
| 454 | msb_sixteen_bit_2 | 0.85030932 | 4093 | 77 |
| 454 | msb_fifteen_bit_2 | 0.85030932 | 4093 | 77 |
| 454 | msb_fourteen_bit_2 | 0.85030932 | 16257 | 77 |
| 454 | msb_thirteen_bit_2 | 0.85030932 | 8065 | 77 |
| 453 | byte_0 | 0.84890293 | 18 | 19 |
| 454 | msb_thirteen_bit_5 | 0.84839341 | 64 | 44 |
| 454 | msb_eleven_bit_5 | 0.84839341 | 64 | 44 |
| 454 | msb_ten_bit_5 | 0.84839341 | 64 | 44 |
| 454 | msb_nine_bit_5 | 0.84839341 | 64 | 44 |
| 454 | byte_6 | 0.84839341 | 254 | 44 |
| 454 | msb_twelve_bit_5 | 0.84839341 | 64 | 44 |
| 454 | msb_fourteen_bit_5 | 0.84839341 | 64 | 44 |
| 170 | msb_sixteen_bit_2 | 0.84165715 | 4521 | 232 |
| 170 | msb_thirteen_bit_2 | 0.84165715 | 8066 | 232 |
| 170 | msb_fourteen_bit_2 | 0.84165715 | 16258 | 232 |
| 170 | msb_fifteen_bit_2 | 0.84165715 | 4521 | 232 |
| 170 | msb_fifteen_bit_5 | 0.84147734 | 4524 | 239 |
| 170 | msb_sixteen_bit_5 | 0.84147734 | 4524 | 239 |
| 170 | msb_thirteen_bit_5 | 0.84147734 | 8064 | 239 |

The time series data obtained from the accelerator pedal calibration recording was analyzed to identify the top results. The highest-ranking channels for acceleration were CAN IDs 453 and 454: specifically channels 453_msb_sixteen_bit_0 and 454_msb_eight_bit_2. The data obtained from these channels, shown in Figure 16, displayed a similarity to the anticipated waveform of the accelerator pedal during the recorded period. However, upon comparison with the expected waveform shown in Figure 15, these CAN channels did not closely match the accelerator pedal calibration test anticipated waveform.

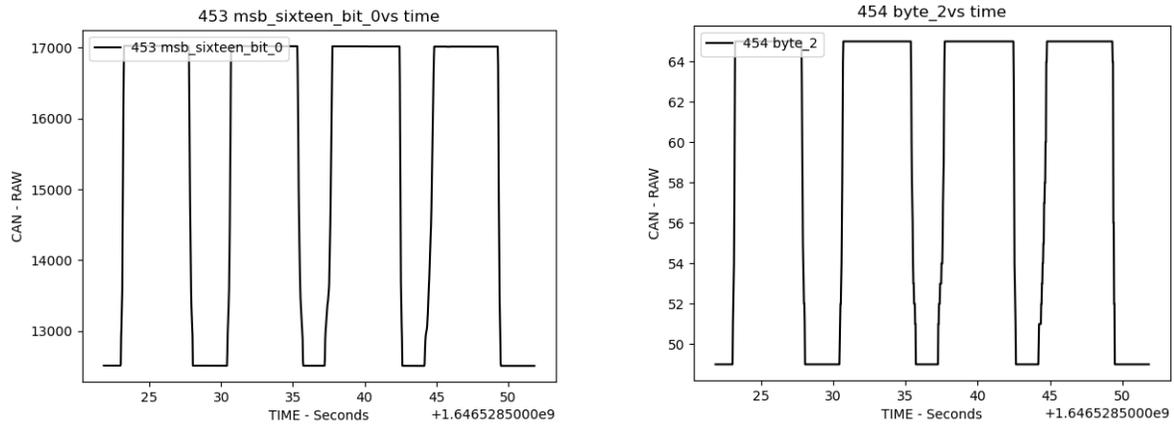

**Figure 16.** Accelerator pedal discovery inconclusive results.

The next part of the algorithm looks for signals that have a smoother range of motion. When an operator moves the accelerator pedal, it is anticipated that the relevant message should have a gradual transition between values and not have significant spikes or unexpected changes. To determine how each signal changes over time, the derivative of every CAN channel's time series data was computed. The standard deviation of each was then calculated, resulting in a single number for each new signal's time series. A final calculation was made which divides the standard deviation of the derivative by the quotient of the range of values recorded. Table 9 presents the data along with new columns presenting the standard deviation of the derivative CAN channel series: ("StDev(*)") and the new smooth ratio ("smooth"). The top 25 channels are included in this table.

**Table 9.** Appending Columns StDev(*) and Smooth to Acceleration Channels.

| ID | Channel | Correlation | Range | Unique | StDev(*) | Smooth |
|---|---|---|---|---|---|---|
| 453 | msb_sixteen_bit_0 | 0.86757076 | 4521 | 126 | 253 | 6 |
| 453 | msb_fifteen_bit_0 | 0.86757076 | 4521 | 126 | 253 | 6 |
| 453 | msb_thirteen_bit_0 | 0.86757076 | 7979 | 126 | 1600 | 21 |
| 453 | msb_fourteen_bit_0 | 0.86757076 | 16171 | 126 | 3361 | 21 |
| 454 | msb_fifteen_bit_6 | 0.85032099 | 32262 | 108 | 2417 | 8 |
| 454 | msb_sixteen_bit_6 | 0.85032099 | 65030 | 108 | 5065 | 8 |
| 454 | msb_sixteen_bit_2 | 0.85030932 | 4093 | 77 | 315 | 8 |
| 454 | msb_fifteen_bit_2 | 0.85030932 | 4093 | 77 | 315 | 8 |
| 454 | msb_fourteen_bit_2 | 0.85030932 | 16257 | 77 | 4238 | 27 |
| 454 | msb_thirteen_bit_2 | 0.85030932 | 8065 | 77 | 1972 | 25 |
| 453 | byte_0 | 0.84890293 | 18 | 19 | 1 | 6 |
| 454 | msb_thirteen_bit_5 | 0.84839341 | 64 | 44 | 5 | 8 |
| 454 | msb_eleven_bit_5 | 0.84839341 | 64 | 44 | 5 | 8 |
| 454 | msb_ten_bit_5 | 0.84839341 | 64 | 44 | 5 | 8 |
| 454 | msb_nine_bit_5 | 0.84839341 | 64 | 44 | 5 | 8 |
| 454 | byte_6 | 0.84839341 | 254 | 44 | 69 | 28 |
| 454 | msb_twelve_bit_5 | 0.84839341 | 64 | 44 | 5 | 8 |

| ID | Channel | Correlation | Range | Unique | StDev(*) | Smooth |
|---|---|---|---|---|---|---|
| 454 | msb_fourteen_bit_5 | 0.84839341 | 64 | 44 | 5 | 8 |
| 170 | msb_sixteen_bit_2 | 0.84165715 | 4521 | 232 | 133 | 3 |
| 170 | msb_thirteen_bit_2 | 0.84165715 | 8066 | 232 | 1309 | 17 |
| 170 | msb_fourteen_bit_2 | 0.84165715 | 16258 | 232 | 2685 | 17 |
| 170 | msb_fifteen_bit_2 | 0.84165715 | 4521 | 232 | 133 | 3 |
| 170 | msb_fifteen_bit_5 | 0.84147734 | 4524 | 239 | 62 | 2 |
| 170 | msb_sixteen_bit_5 | 0.84147734 | 4524 | 239 | 62 | 2 |
| 170 | msb_thirteen_bit_5 | 0.84147734 | 8064 | 239 | 459 | 6 |

The vehicle controls discovery algorithm then sorts the final series data on the "smooth" column in ascending order. The algorithm returns all of the CAN channels that were equal to the smallest increment size displayed in the "smooth" column. The results are shown in Table 10.

Table 10. 2016 GMC Sierra 1500 Potential Accelerator Pedal Channels.

| ID | Channel | Correlation | Range | Unique | StDev(*) | Smooth |
|---|---|---|---|---|---|---|
| 190 | lsb_sixteen_bit_2 | 0.80393274 | 252 | 64 | 1 | 1 |
| 190 | lsb_fifteen_bit_2 | 0.80358291 | 253 | 127 | 2 | 1 |
| 190 | lsb_fourteen_bit_2 | 0.80329673 | 255 | 249 | 2 | 1 |
| 190 | lsb_thirteen_bit_2 | 0.79816976 | 255 | 249 | 2 | 1 |
| 190 | byte_2 | 0.79816976 | 255 | 249 | 2 | 1 |
| 190 | lsb_twelve_bit_2 | 0.79816976 | 255 | 249 | 2 | 1 |
| 190 | lsb_eleven_bit_2 | 0.79816976 | 255 | 249 | 2 | 1 |
| 190 | lsb_ten_bit_2 | 0.79816976 | 255 | 249 | 2 | 1 |
| 190 | lsb_nine_bit_2 | 0.79816976 | 255 | 249 | 2 | 1 |
| 190 | msb_sixteen_bit_2 | 0.79816892 | 65280 | 249 | 347 | 1 |
| 170 | byte_7 | 0.7981634 | 255 | 249 | 2 | 1 |
| 201 | lsb_eleven_bit_4 | 0.7981393 | 254 | 250 | 2 | 1 |
| 201 | byte_4 | 0.7981393 | 254 | 250 | 2 | 1 |
| 201 | lsb_nine_bit_4 | 0.7981393 | 254 | 250 | 2 | 1 |
| 201 | lsb_ten_bit_4 | 0.7981393 | 254 | 250 | 2 | 1 |
| 201 | lsb_twelve_bit_4 | 0.7981393 | 254 | 250 | 2 | 1 |
| 201 | msb_sixteen_bit_4 | 0.7980831 | 65024 | 250 | 345 | 1 |
| 201 | lsb_sixteen_bit_4 | 0.78763181 | 254 | 250 | 2 | 1 |
| 201 | lsb_thirteen_bit_4 | 0.78676656 | 254 | 250 | 2 | 1 |
| 201 | lsb_fifteen_bit_4 | 0.78662138 | 254 | 250 | 2 | 1 |
| 201 | lsb_fourteen_bit_4 | 0.78662138 | 254 | 250 | 2 | 1 |

The time series data during the accelerator pedal calibration recording for these top results was analyzed. Each of the CAN channels outputted from the vehicle controls discovery algorithm appeared to be a direct representation of the accelerator pedal data. Figure 17 compares the CAN channel 190_lsb_sixteen_bit_1 data from the calibration recording for the accelerator pedal (left) to the expected waveform (right).

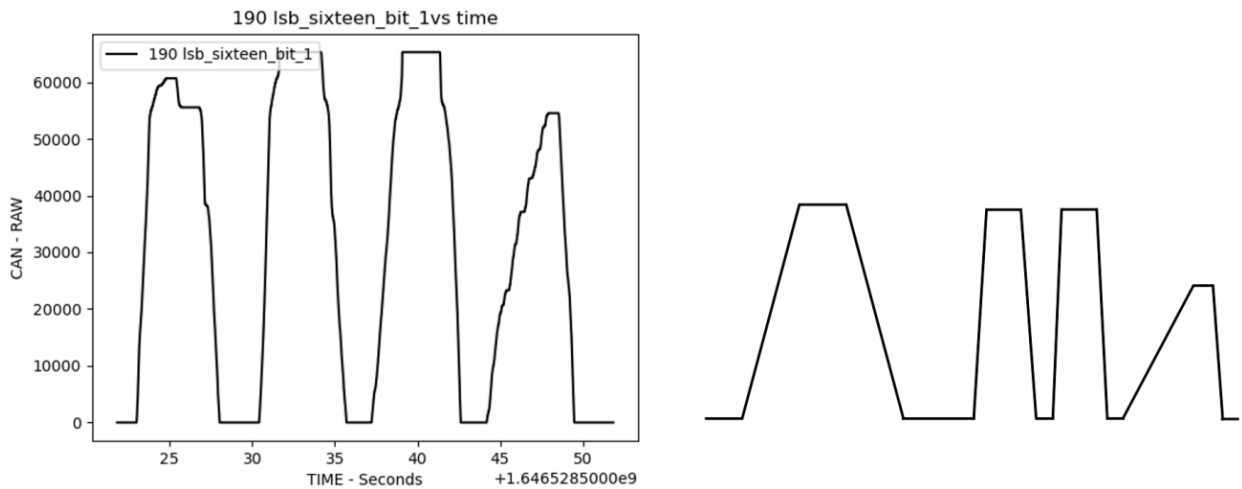

**Figure 17.** Accelerator pedal discovery.

*4.3. Accelerator Pedal Results*

The highest correlated results for CAN channels relating to the accelerator pedal were shown in Table 10. Figure 18 shows data from one of those channels that is clearly related to the accelerator pedal. The top images show data from the CAN channel data during the accelerator pedal calibration recording. The bottom images show the CAN channel data during the vehicle trip recording.

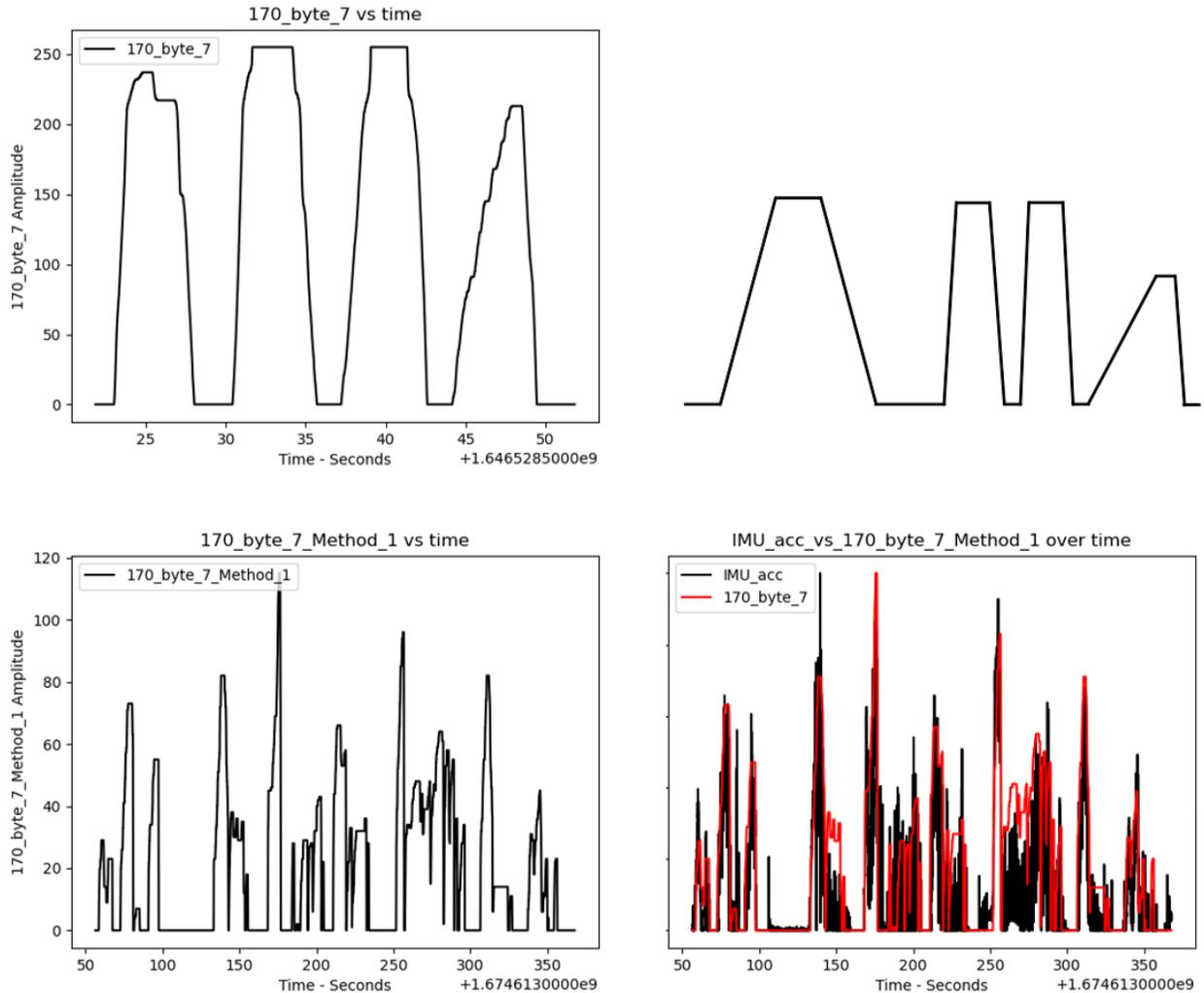
**Figure 18.** Accelerator pedal verification: 170 Sixteen-bit (lsb) channel 6.

*4.4. Deceleration Reverse Engineering*

In this section the reverse engineering results for the brake pedal will be presented. The IMU sensor was arranged within the vehicle in a manner such that accelerations in the Y axis corresponded to the vehicle's forward and aft movements. In this case, vehicle deceleration was represented by positive values measured on the IMU's Y axis. As discussed, in regards to the rate of change correlation algorithm presented in Section 3.5, the Y axis data from the IMU sensor underwent filtering to separate the vehicle's measured deceleration signals as a distinct signal. Figure 19 displays data recorded during a trip with the primary test vehicle, a 2016 GMC Sierra.

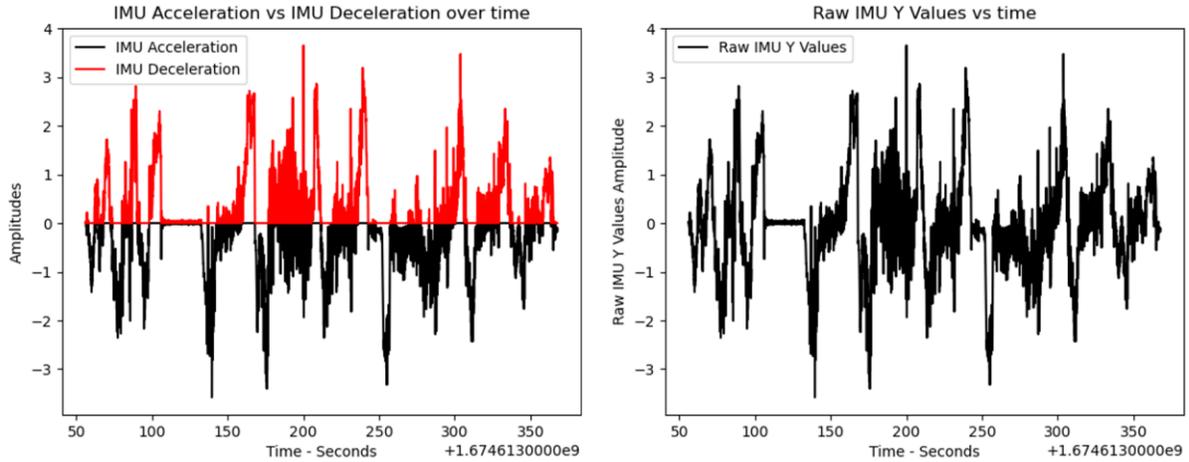

**Figure 19.** IMU deceleration values.

The rate of change correlation algorithm uses the deceleration data as a baseline for correlation with all of the CAN channels recorded during the drive. For every CAN channel, a percent correlation was calculated to determine how closely the specific recorded CAN channel related to vehicle deceleration. The top 25 channel results from this correlation are shown in Table 11.

**Table 11.** Deceleration results.

| ID | Channel | Correlation |
|---|---|---|
| 532 | msb_fourteen_bit_1 | 0.67124106 |
| 761 | msb_thirteen_bit_5 | 0.67116203 |
| 510 | msb_twelve_bit_0 | 0.6684725 |
| 209 | msb_eleven_bit_0 | 0.66658151 |
| 510 | msb_ten_bit_1 | 0.66579049 |
| 508 | msb_eleven_bit_4 | 0.65483176 |
| 761 | msb_fourteen_bit_5 | 0.6521501 |
| 532 | msb_fifteen_bit_1 | 0.65212751 |
| 761 | lsb_eleven_bit_5 | 0.6225587 |
| 510 | byte_0 | 0.62237047 |
| 510 | msb_sixteen_bit_0 | 0.6125464 |
| 510 | msb_fourteen_bit_0 | 0.6125464 |
| 510 | msb_fifteen_bit_0 | 0.6125464 |
| 510 | msb_thirteen_bit_0 | 0.6125464 |
| 209 | msb_thirteen_bit_0 | 0.61191694 |
| 209 | msb_fourteen_bit_0 | 0.61191694 |
| 209 | msb_twelve_bit_0 | 0.61191694 |
| 510 | msb_eleven_bit_1 | 0.61156863 |
| 510 | msb_twelve_bit_1 | 0.61156863 |
| 190 | lsb_twelve_bit_0 | 0.60906623 |
| 532 | msb_sixteen_bit_1 | 0.60672203 |
| 532 | msb_nine_bit_0 | 0.60671468 |
| 532 | byte_1 | 0.60671468 |
| 532 | lsb_nine_bit_1 | 0.60671468 |
| 761 | msb_fifteen_bit_5 | 0.60667765 |

Channels associated with CAN IDs 532 and 761 demonstrated a strong correlation with deceleration. The channel designated for ID 532 was 532_msb_fourteen_bit_1, while the channel associated with ID 761 was 761_msb_thirteen_bit_5. The graphs shown in Figure 20 and Figure 21 display the values of each channel over the course of the drive, in relation to acceleration.

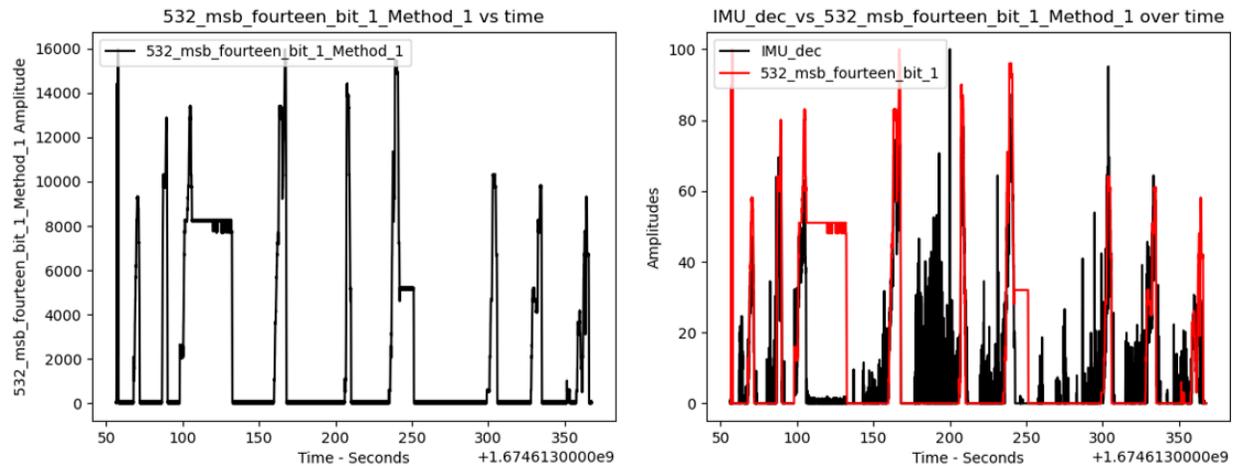

**Figure 20.** Deceleration Results CAN Channel 532_msb_fourteen_bit_1.

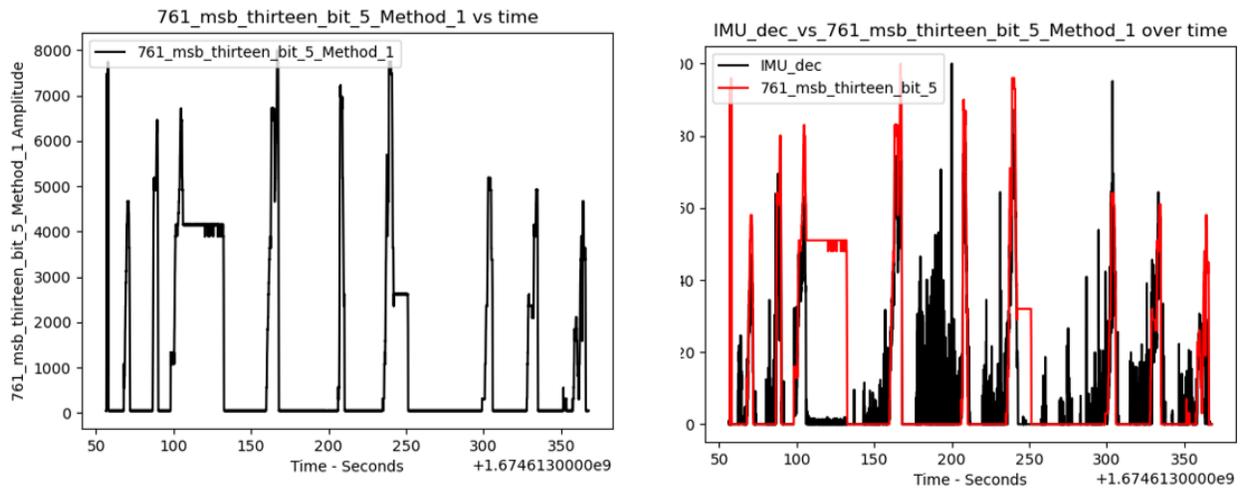

**Figure 21.** Deceleration Results CAN Channel 761_msb_thirteen_bit_5.

*4.5. Brake Pedal Reverse Engineering*

The times series data from the accelerator pedal calibration recording was graphed and is shown in Figure 22. Channel IDs 532 and 761 both had strong correlations with deceleration; however, their values remained constant during the recording of the brake pedal calibration. This, again, shows the importance of the vehicle controls discovery algorithm to identify which of the acceleration CAN channels relate to the brake pedal.

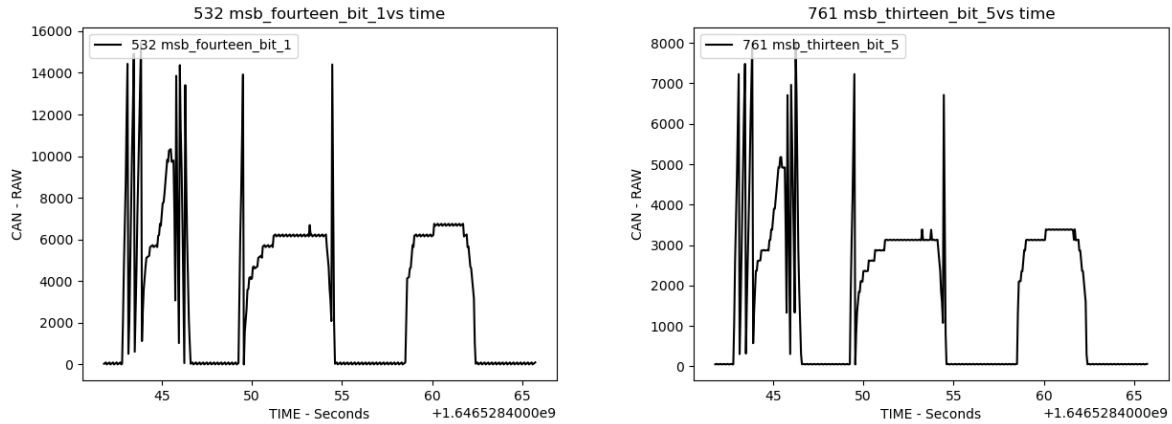

**Figure 22.** Deceleration results during brake calibration recording.

Figure 23 presents the anticipated waveform that should be generated by the brake pedal being depressed and released during its calibration recording.

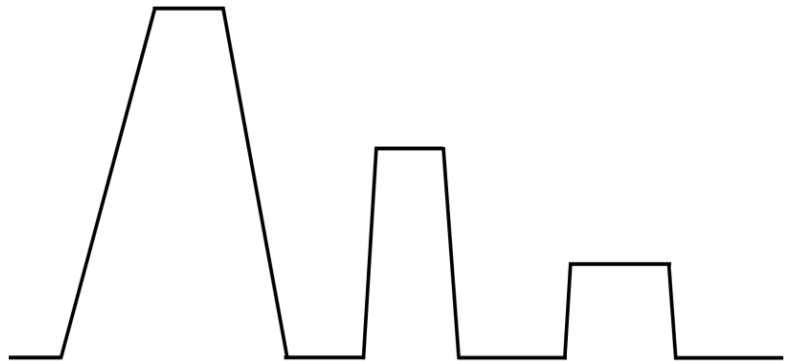

**Figure 23.** Anticipated brake pedal waveform.

The vehicle controls discovery algorithm takes the dataset from the deceleration correlation results and adds additional calculated values for each CAN channel. The algorithm processes the brake pedal recording and identifies the range of values (shown in the range column in Table 12) and number of unique values (shown in the unique column) recorded for each of the CAN channels in the deceleration correlation results. The top 25 of these channels are displayed in Table 12.

**Table 12.** Appending range and unique columns for deceleration channels.

| ID | Channel | Correlation | Range | Unique |
|---|---|---|---|---|
| 532 | msb_fourteen_bit_1 | 0.67124106 | 15425 | 82 |
| 761 | msb_thirteen_bit_5 | 0.67116203 | 7946 | 77 |
| 510 | msb_twelve_bit_0 | 0.6684725 | 3984 | 109 |
| 209 | msb_eleven_bit_0 | 0.66658151 | 2044 | 451 |
| 510 | msb_ten_bit_1 | 0.66579049 | 980 | 92 |
| 508 | msb_eleven_bit_4 | 0.65483176 | 2045 | 250 |
| 761 | msb_fourteen_bit_5 | 0.6521501 | 15626 | 104 |
| 532 | msb_fifteen_bit_1 | 0.65212751 | 31297 | 104 |
| 761 | lsb_eleven_bit_5 | 0.6225587 | 95 | 49 |
| 510 | byte_0 | 0.62237047 | 22 | 23 |
| 510 | msb_sixteen_bit_0 | 0.6125464 | 5722 | 109 |
| 510 | msb_fourteen_bit_0 | 0.6125464 | 5722 | 109 |
| 510 | msb_fifteen_bit_0 | 0.6125464 | 5722 | 109 |
| 510 | msb_thirteen_bit_0 | 0.6125464 | 5722 | 109 |
| 209 | msb_thirteen_bit_0 | 0.61191694 | 5447 | 481 |
| 209 | msb_fourteen_bit_0 | 0.61191694 | 5447 | 481 |
| 209 | msb_twelve_bit_0 | 0.61191694 | 4079 | 452 |
| 510 | msb_eleven_bit_1 | 0.61156863 | 1420 | 98 |
| 510 | msb_twelve_bit_1 | 0.61156863 | 1420 | 98 |
| 190 | lsb_twelve_bit_0 | 0.60906623 | 172 | 44 |
| 532 | msb_sixteen_bit_1 | 0.60672203 | 59489 | 123 |
| 532 | msb_nine_bit_0 | 0.60671468 | 232 | 67 |
| 532 | byte_1 | 0.60671468 | 232 | 67 |
| 532 | lsb_nine_bit_1 | 0.60671468 | 232 | 67 |
| 761 | msb_fifteen_bit_5 | 0.60667765 | 29706 | 122 |

The vehicle controls discovery algorithm then removes the CAN channels that are constant during the brake pedal recording. In this case, the best correlating results had an adequate number of unique values, therefore no changes were made to the brake pedal candidates.

Just like during the discovery of the accelerator pedal CAN channels, the vehicle controls discovery algorithm looks for signals that have a smooth range of motion. Table 13 presents the top 25 brake pedal CAN channels, with new columns presenting the standard deviation of the derivative CAN channel series ("StDev(*)") and the smooth ratio ("smooth"), which is a percent.

Table 13. Appending columns *StDev(*)* and *Smooth* to deceleration channels.

| ID | Channel | Correlation | Range | Unique | StDev(*) | Smooth |
|---|---|---|---|---|---|---|
| 532 | msb_fourteen_bit_1 | 0.67124106 | 15425 | 82 | 1875 | 13 |
| 761 | msb_thirteen_bit_5 | 0.67116203 | 7946 | 77 | 942 | 12 |
| 510 | msb_twelve_bit_0 | 0.6684725 | 3984 | 109 | 403 | 11 |
| 209 | msb_eleven_bit_0 | 0.66658151 | 2044 | 451 | 214 | 11 |
| 510 | msb_ten_bit_1 | 0.66579049 | 980 | 92 | 109 | 12 |
| 508 | msb_eleven_bit_4 | 0.65483176 | 2045 | 250 | 273 | 14 |
| 761 | msb_fourteen_bit_5 | 0.6521501 | 15626 | 104 | 1075 | 7 |
| 532 | msb_fifteen_bit_1 | 0.65212751 | 31297 | 104 | 2139 | 7 |
| 761 | lsb_eleven_bit_5 | 0.6225587 | 95 | 49 | 8 | 9 |
| 510 | byte_0 | 0.62237047 | 22 | 23 | 1 | 5 |

| 510 | msb_sixteen_bit_0 | 0.6125464 | 5722 | 109 | 124 | 3 |
|---|---|---|---|---|---|---|
| 510 | msb_fourteen_bit_0 | 0.6125464 | 5722 | 109 | 124 | 3 |
| 510 | msb_fifteen_bit_0 | 0.6125464 | 5722 | 109 | 124 | 3 |
| 510 | msb_thirteen_bit_0 | 0.6125464 | 5722 | 109 | 124 | 3 |
| 209 | msb_thirteen_bit_0 | 0.61191694 | 5447 | 481 | 26 | 1 |
| 209 | msb_fourteen_bit_0 | 0.61191694 | 5447 | 481 | 26 | 1 |
| 209 | msb_twelve_bit_0 | 0.61191694 | 4079 | 452 | 216 | 6 |
| 510 | msb_eleven_bit_1 | 0.61156863 | 1420 | 98 | 32 | 3 |
| 510 | msb_twelve_bit_1 | 0.61156863 | 1420 | 98 | 32 | 3 |
| 190 | lsb_twelve_bit_0 | 0.60906623 | 172 | 44 | 4 | 3 |
| 532 | msb_sixteen_bit_1 | 0.60672203 | 59489 | 123 | 941 | 2 |
| 532 | msb_nine_bit_0 | 0.60671468 | 232 | 67 | 6 | 3 |
| 532 | byte_1 | 0.60671468 | 232 | 67 | 6 | 3 |
| 532 | lsb_nine_bit_1 | 0.60671468 | 232 | 67 | 6 | 3 |
| 761 | msb_fifteen_bit_5 | 0.60667765 | 29706 | 122 | 480 | 2 |

The vehicle controls discovery algorithm then sorts the data by the smooth column values, in ascending order. The algorithm returns all of the CAN channels that were equal to the smallest increment size displayed in the smooth column. The results are shown in Table 14. For the channels in Table 14, a smooth value of 1 implies that CAN channel data changes at 1% of its possible range of values.

Table 14. 2016 GMC Sierra 1500 potential brake pedal channels.

| ID | Channel | Correlation | Range | Unique | StDev(*) | Smooth |
|---|---|---|---|---|---|---|
| 209 | msb_fourteen_bit_0 | 0.61191694 | 5447 | 481 | 26 | 1 |
| 209 | msb_thirteen_bit_0 | 0.61191694 | 5447 | 481 | 26 | 1 |
| 190 | lsb_sixteen_bit_0 | 0.59129758 | 42508 | 368 | 253 | 1 |
| 190 | msb_sixteen_bit_1 | 0.56810696 | 42496 | 157 | 315 | 1 |
| 241 | lsb_thirteen_bit_1 | 0.55803216 | 129 | 128 | 1 | 1 |
| 241 | byte_1 | 0.55803216 | 129 | 128 | 1 | 1 |
| 241 | lsb_nine_bit_1 | 0.55803216 | 129 | 128 | 1 | 1 |
| 241 | lsb_ten_bit_1 | 0.55803216 | 129 | 128 | 1 | 1 |
| 241 | lsb_eleven_bit_1 | 0.55803216 | 129 | 128 | 1 | 1 |
| 241 | lsb_twelve_bit_1 | 0.55803216 | 129 | 128 | 1 | 1 |
| 241 | msb_nine_bit_0 | 0.55803216 | 129 | 128 | 1 | 1 |
| 241 | msb_sixteen_bit_1 | 0.55797043 | 33028 | 356 | 186 | 1 |

The time series data during the brake pedal calibration recording for these top results was analyzed. Each of the CAN channels outputted from the vehicle controls discovery algorithm appeared to be a direct representation of the brake pedal data. Figure 24 displays the CAN channel 209_msb_fourteen_bit_0 (left) and CAN channel 241_lsb_sixteen_bit_0 (right) data from the calibration recording for the brake pedal.

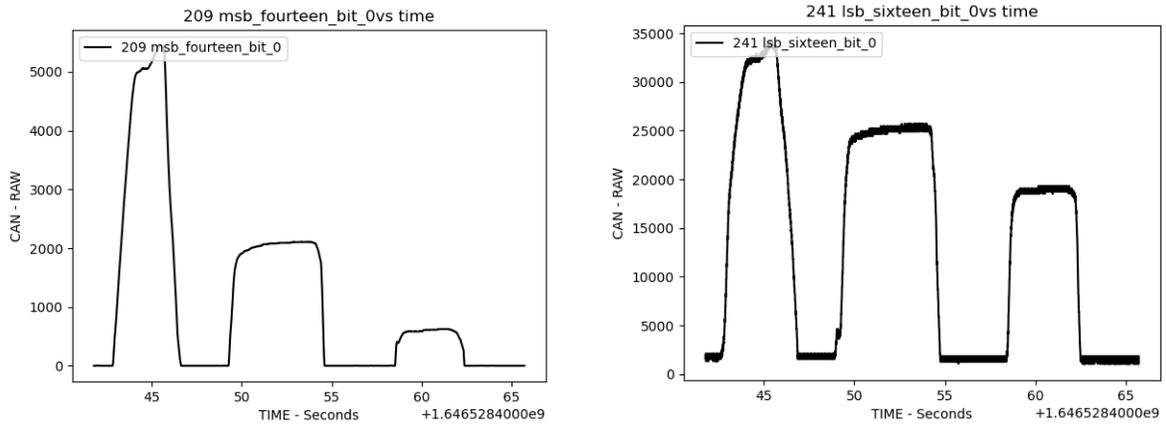

**Figure 24.** Brake Pedal Channels During Calibration.

*4.6. Brake Pedal Results*

The best correlated results for CAN channels relating to the brake pedal were identified. Figure 25 shows one of the channels that is clearly related to the brake pedal. The top images show the CAN channel data during the brake pedal calibration recording. The bottom images show the CAN channel data during the vehicle trip recording.

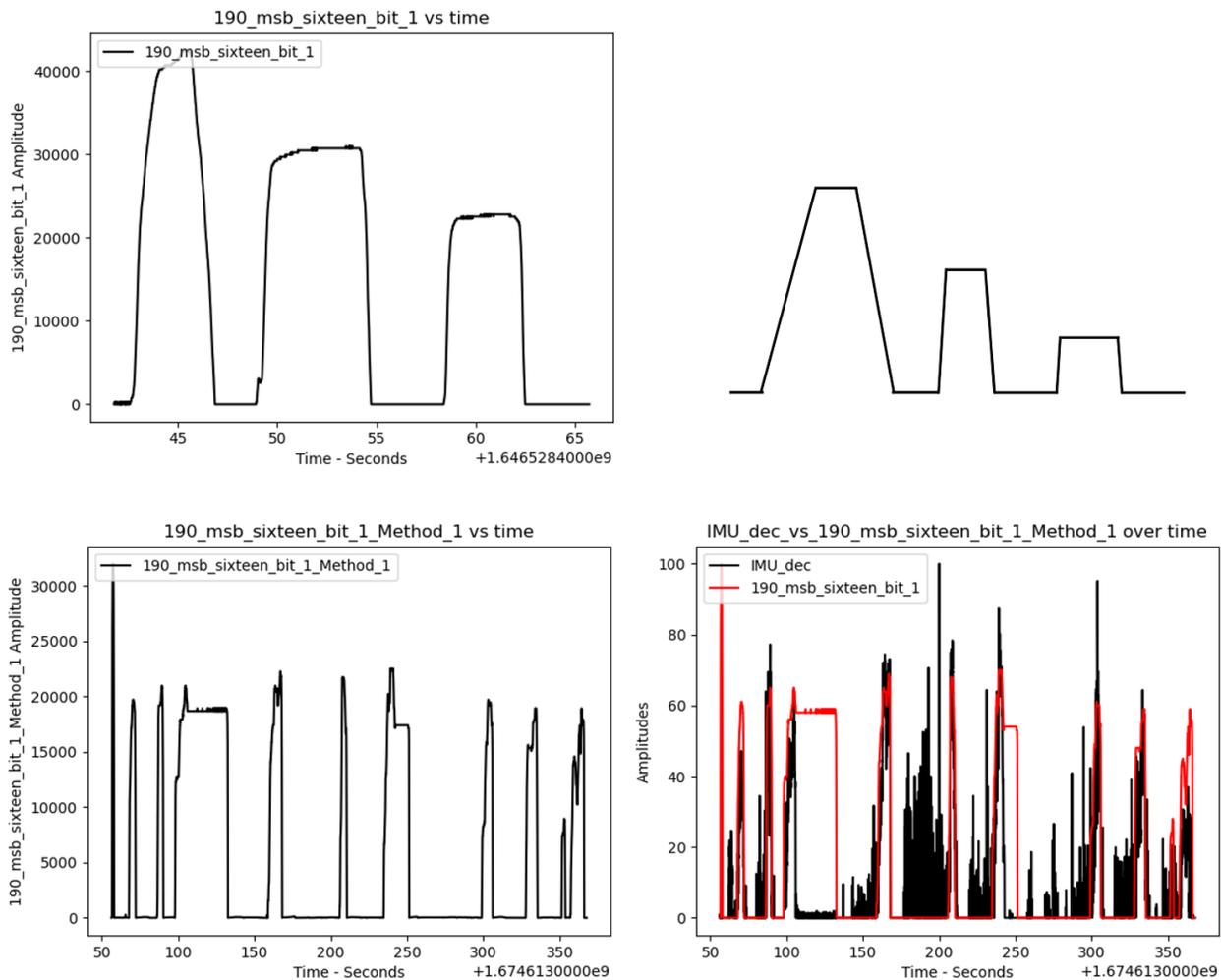
**Figure 25.** Brake pedal verification: 190 Sixteen-bit (msb) channel 1.

## 5. Discussion

This section reviews and evaluates the results from the testing discussed previously in this paper. First, an interpretation of results is presented. This is followed by a comparison to previous research in this area. This section concludes with a discussion of the limitations and recommendations for areas of future research.

### 5.1. Interpretation and Analysis of Results

This paper has shown that it's feasible to reverse engineer CAN messages pertaining to vehicle controls without any prior knowledge of the vehicle's CAN format. The system proposed in this study successfully identified the CAN channels for the accelerator pedal and the brake pedal of each vehicle tested.

The correlation performance between the acceleration and deceleration were slightly different. It was found that the deceleration correlation was slightly weaker than acceleration correlation, on average, across all vehicles tested. One possible reason for this is the time where the vehicle was at rest and the driver was still potentially holding down the brake pedal. In these time frames, the vehicles velocity was already at zero, and therefore the vehicle's speed could not decelerate any further. It is possible that these times diminished the correlation values between the CAN messages for the brake pedal and the IMU deceleration. Regardless, the results still show that the method presented in this paper could serve as a reverse engineering tool for multiple vehicles from different manufacturers.

5.1.1. Correlating CAN Channels to Vehicle Actions

In general, for the categories of acceleration and deceleration, there were multiple CAN channels that were shown to have a strong correlation. It was also shown that there was no overlap of strongly correlated CAN channels between acceleration and deceleration. Interestingly, the top values that strongly related to acceleration did not strongly relate to the accelerator pedal itself. This seems logical, given the fact that the accelerator pedal is merely an enabler that starts a chain reaction of events that ultimately leads to the vehicle moving forward.

5.1.2. Identifying CAN Channels Relating to Vehicle Controls

The proposed algorithm was shown to be effective for identifying which CAN channels related to the vehicle's accelerator pedal and brake pedal, using the collected lists of CAN channels that related to vehicle acceleration and deceleration. The results are a success since applicable CAN channels were found for each vehicle tested. The algorithm was shown to perform similarly across all tested vehicles. The specific CAN channels identified were found to differ between the different manufacturers; however, there were multiple vehicles analyzed form the GM/Chevy product line and their CAN channel results were very similar.

## *5.2. Comparisons to Previous Research*

The results presented in this paper align with Pese, et al.'s [20] analysis regarding the difficulty of quantifying performance, due to the differences between the CAN systems of different vehicle OEMs. Pese, et al. also had access to vehicle DBC files [46], provided by an OEM to use as ground truth values to analyze the performance of their system. This experiment did not have access to those DBC files. Therefore, the method of visual verification of the anticipated waveform of the signal during the CAN channel's calibration recording and the CAN channel's data during vehicle driving recordings was used. This research did not use CAN injection to verify CAN messages, unlike some other papers.

The work presented herein focused on the automated discovery of the CAN channels for the vehicle's accelerator pedal and brake pedal. Like some previous studies, multiple CAN channels were found for each vehicle. The results for the 2016 GMC Sierra were presented in the previous section and results for several other vehicles are included in Appendix A.

Unlike the methods used in previous studies, the proposed system does not require information about the vehicle or OBD-II standards. It can achieve similar results by simply observing, making it less intrusive and potentially more resilient to changes in vehicle or industry CAN bus standards. An abstraction layer could prospectively be added to make it useable with other similar serial interfaces.

## *5.3. Limitations and Recommendations for Future Research*

One limitation of the approach presented in this study is the number of possible combinations of CAN channels that are generated and assessed for each vehicle tested. Since there was no software (like READ [10], LibreCAN [20], or CANMatch [5]) developed to identify the signal boundaries within the CAN messages, the proposed system uses an exhaustive search technique to identify numerous CAN channel combinations. In many cases, this returned results where different bit lengths were derived from the same

CAN frame ID and endianness, as shown in Table 10, which presented the results for the 2016 GMC Sierra 1500 accelerator pedal channels. This same issue is also present in the data from other vehicles tested, presented in Appendix A. An additional method for narrowing down these possibilities may be beneficial to enhance system performance speed.

Since the algorithms are not meant to be used in real-time, the time required to converge on an outcome was acceptable for use during post processing. Limiting the number of possibilities of channels to assess could be a goal of future work. This could reduce the time required for CAN channel identification and it would also reduce the time and processing power required for the algorithms later in the process.

Another area for potential improvement is in the IMU. Accelerometer measurements in IMU's are known to be noisy. Assuming the CAN messages relating to the accelerator pedal and brake pedals do not have similar noise, this noise can cause discrepancies in the correlation between the two. Improvements to the IMU, particularly by filtering, could improve the correlation strengths, and therefore could provide greater dispersion between CAN channels that only indirectly relate to the accelerator and brake pedals, and those that directly relate to the accelerator and brake pedals.

Overall, the results presented show that the proposed method could serve as a useful reverse engineering tool for multiple vehicles from various OEMs.

## 6. Conclusions

This paper has presented a method for reverse engineering that can be used to identify CAN channels related to the accelerator and brake pedals of a vehicle without any prior knowledge of the vehicle's CAN bus format. The system developed for this research included an IMU data collection module and a CAN data collection module. The data from these interfaces was recorded in parallel during vehicle operation. After recording, two software algorithms were used to process this data to identify channels associated with the accelerator and brake pedals. The first algorithm identified the specific CAN channels related to vehicle acceleration and deceleration through preprocessing and correlation analysis. The second algorithm identified which CAN channels related to the accelerator and brake pedals using stationary recordings collected for each vehicle control.

The results from testing with multiple vehicles showed that this method was effective in identifying the CAN channels related to each vehicle's accelerator and brake pedals. However, the sample size of vehicles tested was small, compared to the total number of vehicles and manufacturers in the marketplace, and the exhaustive search approach for identifying CAN channels limits the system from being used for real-time applications. Despite the limitations, this work demonstrated a clear potential for developing aftermarket autonomous vehicle kits. This same information and techniques could also be useful for identifying other vehicle controls and to hackers and cyber defenders alike.

Notably, the proposed solution provides the ability to rapidly adapt to CAN standards or messages changing. Additionally, the system could be abstracted so that it could be used with multiple serial communication protocols.

This paper has demonstrated a method for autonomous CAN reverse engineering, using post-processing algorithms. As real-time reverse engineering may be useful, in some cases, future work can focus on developing an algorithm suitable for this.

## References

[1]     Tesla, "tesla.com/model3."

[2]   "comma – introducing comma three." https://comma.ai/ (accessed Aug. 10, 2022).
[3]   A. Goyal and A. Thakur, "An Overview of Drive by Wire Technology for Automobiles," in *2019 International Conference on Automation, Computational and Technology Management (ICACTM)*, 2019, pp. 108–110. doi: 10.1109/ICACTM.2019.8776712.
[4]   C. Rossi, A. Tilli, and A. Tonielli, "Robust control of a throttle body for drive by wire operation of automotive engines," *IEEE Transactions on Control Systems Technology*, vol. 8, no. 6, 2000, doi: 10.1109/87.880604.
[5]   A. Buscemi, I. Turcanu, G. Castignani, R. Crunelle, and T. Engel, "CANMatch: A Fully Automated Tool for CAN Bus Reverse Engineering Based on Frame Matching," *IEEE Trans Veh Technol*, vol. 70, no. 12, 2021, doi: 10.1109/TVT.2021.3124550.
[6]   "comma - Compatibility." https://comma.ai/vehicles (accessed Aug. 10, 2022).
[7]   W. Xing, H. Chen, and H. Ding, "The application of controller area network on vehicle," in *Proceedings of the IEEE International Vehicle Electronics Conference (IVEC'99) (Cat. No.99EX257)*, 1999, pp. 455–458 vol.1. doi: 10.1109/IVEC.1999.830728.
[8]   K. Johansson, M. Törngren, and L. Nielsen, "Vehicle Applications of Controller Area Network," 2005, pp. 741–765. doi: 10.1007/0-8176-4404-0_32.
[9]   R. Isermann, R. Schwarz, and S. Stolzl, "Fault-tolerant drive-by-wire systems," *IEEE Control Systems Magazine*, vol. 22, no. 5, pp. 64–81, 2002, doi: 10.1109/MCS.2002.1035218.
[10]  M. Marchetti and D. Stabili, "READ: Reverse engineering of automotive data frames," *IEEE Transactions on Information Forensics and Security*, vol. 14, no. 4, 2019, doi: 10.1109/TIFS.2018.2870826.
[11]  C. Miller and C. Valasek, "A Survey of Remote Automotive Attack Surfaces," *Defcon 22*, 2014.
[12]  C. Valasek and C. Miller, "Adventures in Automotive Networks and Control Units," *Technical White Paper*, 2013.
[13]  C. Miller and C. Valasek, "Remote Exploitation of an Unaltered Passenger Vehicle," *Defcon 23*, vol. 2015, 2015.
[14]  D. Spaar, "Beemer, Open Thyself! – Security vulnerabilities in BMW's ConnectedDrive," *c't Magazin*, 2015.
[15]  "DARPA Hacked a Chevy Impala Through Its OnStar System - autoevolution." https://www.autoevolution.com/news/darpa-hacked-a-chevy-impala-through-its-onstar-system-video-92194.html (accessed Aug. 10, 2022).
[16]  "Tesla car doors can be hacked." https://money.cnn.com/2014/03/31/technology/security/tesla-hack/ (accessed Aug. 10, 2022).
[17]  "Tesla Responds to Chinese Hack With a Major Security Upgrade | WIRED." https://www.wired.com/2016/09/tesla-responds-chinese-hack-major-security-upgrade/ (accessed Aug. 10, 2022).
[18]  T. Huybrechts, Y. Vanommeslaeghe, D. Blontrock, G. Van Barel, and P. Hellinckx, "Automatic reverse engineering of can bus data using machine learning techniques," in *Lecture Notes on Data Engineering and Communications Technologies*, 2018. doi: 10.1007/978-3-319-69835-9_71.
[19]  T. U. Kang, H. M. Song, S. Jeong, and H. K. Kim, "Automated Reverse Engineering and Attack for CAN Using OBD-II," in *IEEE Vehicular Technology Conference*, 2018. doi: 10.1109/VTCFall.2018.8690781.
[20]  M. D. Pesé, T. Stacer, C. Andrés Campos, E. Newberry, D. Chen, and K. G. Shin, "LibreCan: Automated can message translator," in *Proceedings of the ACM Conference on Computer and Communications Security*, Association for Computing Machinery, Nov. 2019, pp. 2283–2300. doi: 10.1145/3319535.3363190.
[21]  "Torque Pro (OBD 2 & Car) - Apps on Google Play." https://play.google.com/store/apps/details?id=org.prowl.torque&hl=en_US&gl=US (accessed Aug. 10, 2022).
[22]  B. Blaauwendraad and V. Kieberl, "Automated reverse-engineering of CAN messages using OBD-II and correlation coefficients," 2020. Accessed: Jun. 10, 2022. [Online]. Available:

[23] N. Navet, Y. Song, F. Simonot-Lion, and C. Wilwert, "Trends in Automotive Communication Systems," *Proceedings of the IEEE*, vol. 93, no. 6, pp. 1204–1223, 2005, doi: 10.1109/JPROC.2005.849725.

[24] Y. Onuma, Y. Terashima, and R. Kiyohara, "ECU software updating in future vehicle networks," in *Proceedings - 31st IEEE International Conference on Advanced Information Networking and Applications Workshops, WAINA 2017*, 2017. doi: 10.1109/WAINA.2017.45.

[25] P. E. Lanigan, S. Kavulya, P. Narasimhan, T. E. Fuhrman, and M. a. Salman, "Diagnosis in Automotive Systems: A Survey," *Current*, 2011.

[26] R. Bosch, "CAN Specification Version 2.0," *Rober Bousch GmbH, Postfach*, vol. 300240, 1991.

[27] C. Hanxing and T. Jun, "Research on the controller area network," in *Proceedings - 2009 International Conference on Networking and Digital Society, ICNDS 2009*, 2009, pp. 251–254. doi: 10.1109/ICNDS.2009.142.

[28] National Instruments, "National Instruments CAN Overview," *https://www.ni.com/en-us/innovations/white-papers/06/controller-area-network--can--overview.html*.

[29] Open System's Interconnection (OSI) Model, "'X.225 : Information technology – Open Systems Interconnection – Connection-oriented Session protocol: Protocol specification,'" *https://web.archive.org/web/20210201064044/https://www.itu.int/rec/T-REC-X.225-199511-I/en*.

[30] www.iso.org, "ISO 11898-1:2015 Road vehicles — Controller area network (CAN) — Part 1: Data link layer and physical signalling," *https://www.iso.org/standard/63648.html*.

[31] S. Corrigan, "Controller Area Network Physical Layer Requirements," *Texas Instruments Application Report , SLLA270-January*, pp. 1–15, 2008.

[32] www.iso.org, "ISO 11898-2:2016 Road vehicles — Controller area network (CAN) — Part 2: High-speed medium access unit," *https://www.iso.org/standard/67244.html*.

[33] SAE International, "High-Speed CAN (HSC) for Vehicle Applications at 500 KBPS J2284/3_201611," *https://www.sae.org/standards/content/j2284/3_201611/*, 2023.

[34] S. Corrigan and I. Interface, "Introduction to the Controller Area Network ( CAN )," *Texas Instruments*, no. August 2002, pp. 1–17, 2016.

[35] "CAN BUS ANALYZER TOOL | Microchip Technology." https://www.microchip.com/en-us/development-tool/APGDT002 (accessed Aug. 10, 2022).

[36] Yahboom, "Yahboom IMU 10-Axis Inertial Navigation ARHS Sensor Module with Accelerometer Gyroscope Magnetometer Barometer Air pressure gauge," *https://category.yahboom.net/products/imu*, 2023.

[37] yahboom, "Welcome to 10-axis IMU Navigation Module repository," *http://www.yahboom.net/study/IMU*.

[38] rkollataj, "Linux kernel driver for Microchip CAN BUS Analyzer Tool," *https://github.com/rkollataj/mcba_usb*, 2017.

[39] Linux Kernel Organization, "SocketCAN - Controller Area Network," *https://www.kernel.org/doc/html/latest/networking/can.html*.

[40] http://wiki.ros.org/socketcan_bridge, "socketcan_bridge," *Wanders, Ivor*, 2016.

[41] "Terminator A Linux Terminal Emulator With Multiple Terminals In One Window." https://www.linuxandubuntu.com/home/terminator-a-linux-terminal-emulator-with-multiple-terminals-in-one-window (accessed Aug. 10, 2022).

[42] "How to Install and Use Terminator Terminal in Linux." https://www.linuxshelltips.com/terminator-terminal-emulator/ (accessed Aug. 10, 2022).

[43] "roslaunch - ROS Wiki." http://wiki.ros.org/roslaunch (accessed Aug. 10, 2022).

[44] The SciPy community, "scipy.stats.pearsonr," *https://docs.scipy.org/doc/scipy/reference/generated/scipy.stats.pearsonr.html*, 2023.

[45] W. McKinney, "pandas: a Foundational Python Library for Data Analysis and Statistics," 2011.

[46]   Automotive Security Research Group, "LibreCAN: Automated CAN Message Translator," *https://www.youtube.com/watch?v=vWSS2G1RBY4*, 2020.

**Appendix A**

In this section, the accelerator and brake pedal data and results are presented for several additional vehicles: a 2016 GMC Sierra 1500, a 2021 GMC Sierra 2500, a 2022 Chevrolet Traverse, a 2009 Chevrolet Impala, a 2006 Volvo XC90, and a 2016 Ford Fusion.

*A.1. 2016 GMC Sierra 1500 Results*

The results for the 2016 GMC Sierra were presented in Section 5. This section includes additional data that supports the results presented previously. Figure A1 presents data for CAN channel 190_byte_2. The top images show the CAN channel actual and projected data from during the accelerator pedal calibration recording. The bottom images show the CAN channel data during the vehicle trip recording.

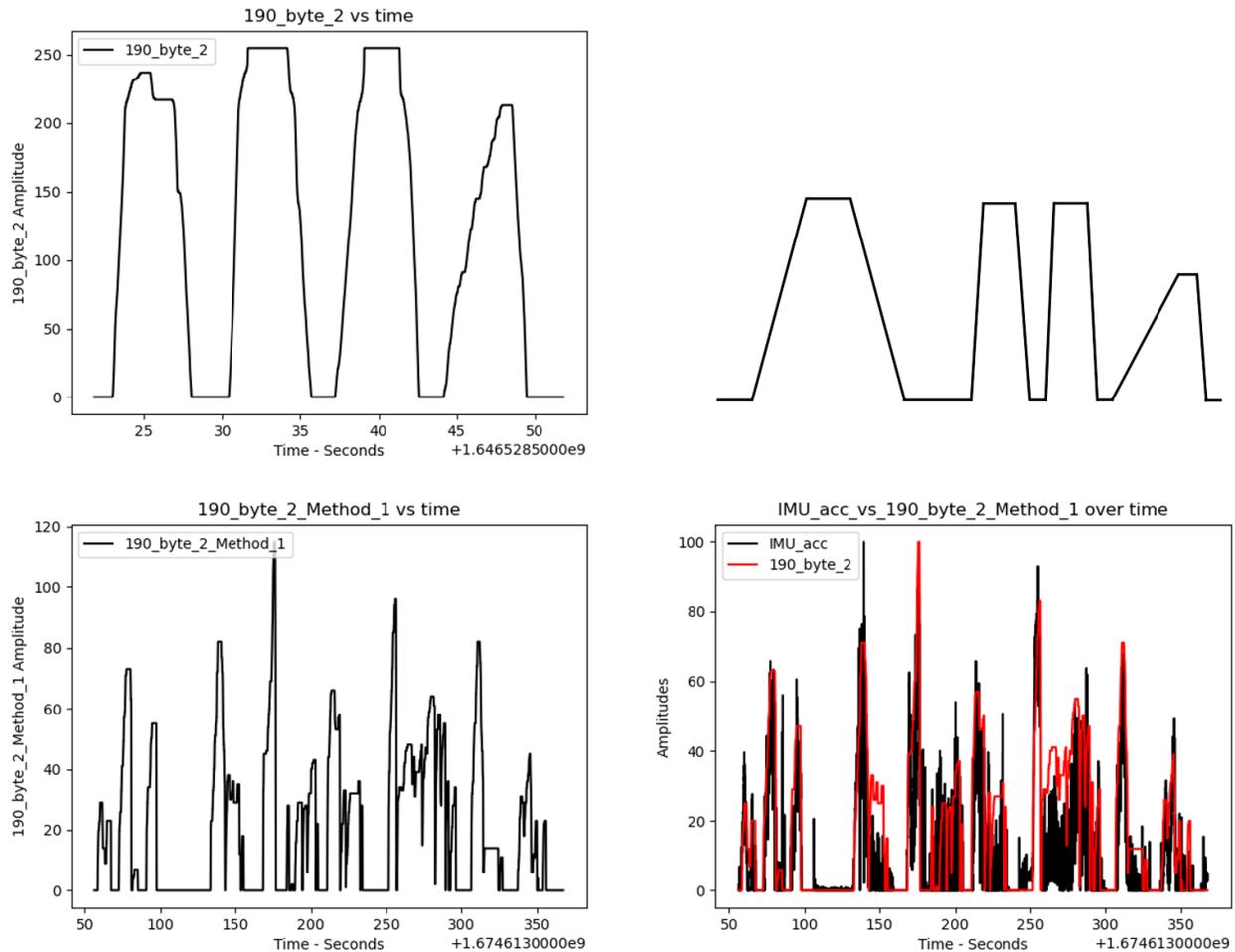

**Figure A1.** 2016 GMC Sierra 1500 Accelerator Pedal Verification: 201 Sixteen-bit (lsb) channel 1.

Figure A2 presents data for CAN channel 241_byte_1. The top images show the CAN channel actual and projected data from during the brake pedal calibration recording. The bottom images show the CAN channel data from during the vehicle trip recording.

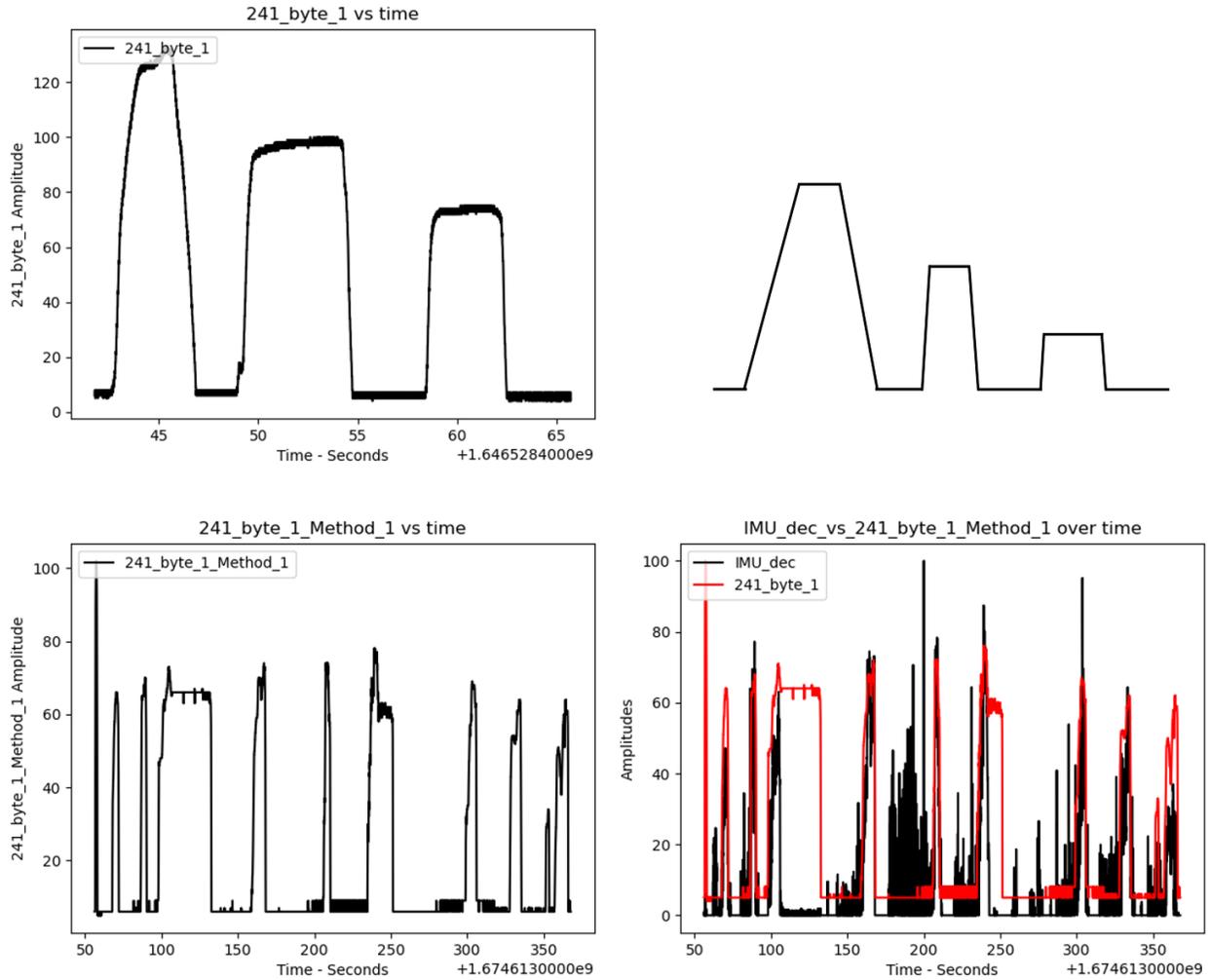

**Figure A2.** 2016 GMC Sierra 1500 Brake Pedal Verification: 241 Eight-bit (msb) Channel 1.

*A.2. 2021 GMC Sierra 2500 Results*

This section presents data for the 2021 GMC Sierra 2500. Table A1 lists the top 25 potential CAN channels for the accelerator pedal. Figure A3 presents data for CAN channel 190_lsb_sixteen_bit_1 and Figure A4 presents data for CAN channel 201_byte_4. For each figure, the top images show the actual and projected CAN channel data during the accelerator pedal calibration recording. The bottom images show the CAN channel data during the vehicle trip recording.

**Table A1.** 2021 GMC Sierra 2500 Accelerator Pedal Results.

| ID | Channel | Correlation | Range | Unique | StDev(*) | Smooth |
|----|---------|-------------|-------|--------|----------|--------|
| 190 | lsb_fifteen_bit_2 | 0.81984312 | 253 | 110 | 2 | 1 |
| 190 | lsb_fourteen_bit_2 | 0.81982617 | 255 | 147 | 2 | 1 |
| 190 | lsb_sixteen_bit_1 | 0.81947624 | 65280 | 147 | 373 | 1 |
| 401 | lsb_sixteen_bit_5 | 0.81944348 | 65024 | 155 | 372 | 1 |
| 201 | msb_sixteen_bit_4 | 0.81941171 | 65024 | 154 | 371 | 1 |
| 201 | byte_4 | 0.81940887 | 254 | 154 | 2 | 1 |

| 201 | lsb_ten_bit_4 | 0.81940887 | 254 | 154 | 2 | 1 |
| 201 | lsb_nine_bit_4 | 0.81940887 | 254 | 154 | 2 | 1 |
| 201 | lsb_eleven_bit_4 | 0.81940887 | 254 | 154 | 2 | 1 |
| 201 | lsb_sixteen_bit_3 | 0.81940853 | 65037 | 189 | 490 | 1 |
| 190 | byte_2 | 0.81932714 | 255 | 147 | 2 | 1 |
| 190 | lsb_thirteen_bit_2 | 0.81932714 | 255 | 147 | 2 | 1 |
| 190 | lsb_nine_bit_2 | 0.81932714 | 255 | 147 | 2 | 1 |
| 190 | lsb_ten_bit_2 | 0.81932714 | 255 | 147 | 2 | 1 |
| 170 | byte_7 | 0.81932714 | 255 | 147 | 2 | 1 |
| 190 | lsb_eleven_bit_2 | 0.81932714 | 255 | 147 | 2 | 1 |
| 190 | lsb_twelve_bit_2 | 0.81932714 | 255 | 147 | 2 | 1 |
| 190 | msb_sixteen_bit_2 | 0.81932711 | 65280 | 147 | 373 | 1 |
| 401 | lsb_nine_bit_6 | 0.8192887 | 254 | 155 | 2 | 1 |
| 401 | msb_sixteen_bit_6 | 0.8192887 | 65278 | 155 | 374 | 1 |
| 401 | lsb_eleven_bit_6 | 0.8192887 | 254 | 155 | 2 | 1 |
| 401 | lsb_fifteen_bit_6 | 0.8192887 | 254 | 155 | 2 | 1 |
| 401 | byte_7 | 0.8192887 | 254 | 155 | 2 | 1 |
| 401 | lsb_fourteen_bit_6 | 0.8192887 | 254 | 155 | 2 | 1 |
| 401 | byte_6 | 0.8192887 | 254 | 155 | 2 | 1 |

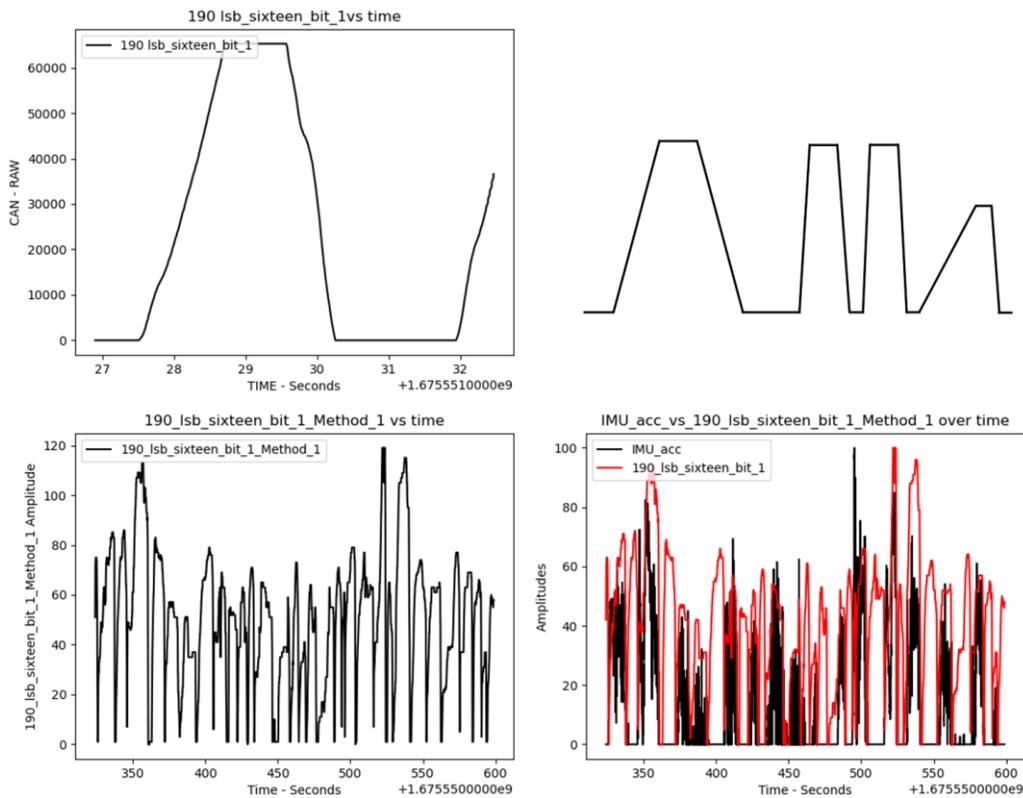

**Figure A3.** 2021 GMC Sierra Accelerator Pedal Channel 190_lsb_sixteen_bit_1.**Figure**

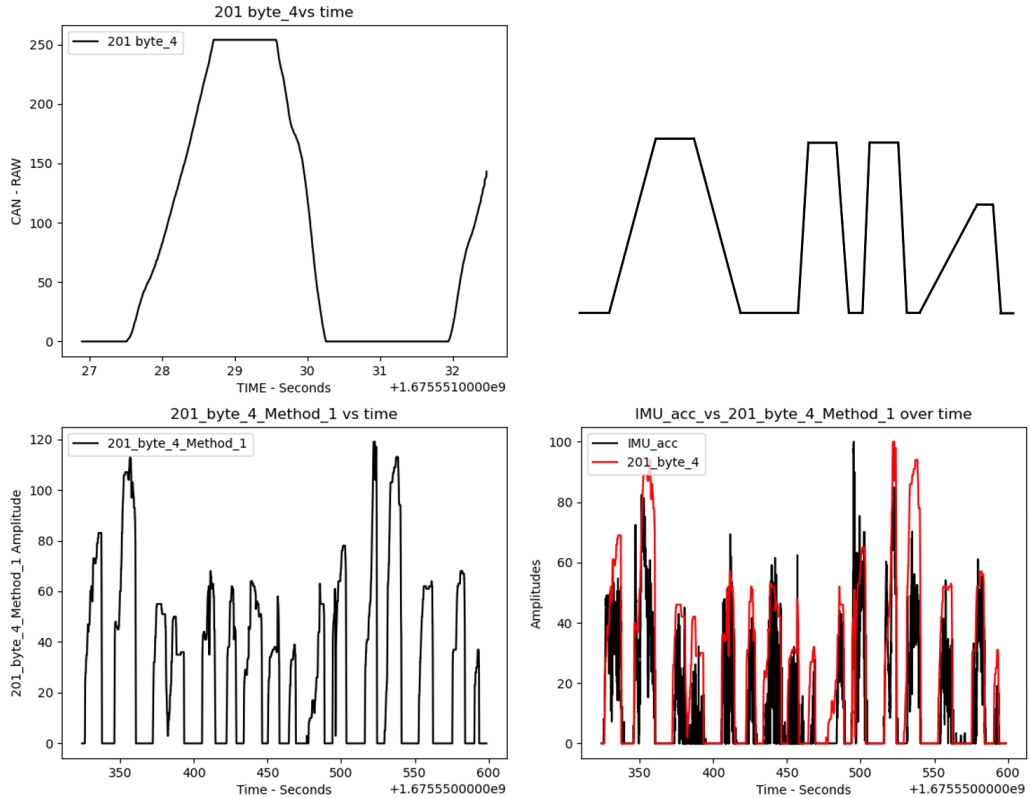
**Figure A4.** 2021 GMC Sierra Accelerator Pedal Channel 201_byte_4.

Table A2 lists the top 25 potential CAN channels for the brake pedal. Figure A5 presents data for CAN channel 241_byte_1 and Figure A6 presents data for CAN channel 241_byte_4. For each figure, the top images show the actual and projected CAN channel data during the brake pedal calibration recording. The bottom images show the CAN channel data during the vehicle trip recording.

**Table A2.** 2021 GMC Sierra 2500 Brake Pedal Results.

| ID | Channel | Correlation | Range | Unique | StDev(*) | Smooth |
|---|---|---|---|---|---|---|
| 190 | lsb_sixteen_bit_0 | 0.868084603 | 27660 | 321 | 243 | 1 |
| 190 | msb_fifteen_bit_1 | 0.867804824 | 27648 | 108 | 264 | 1 |
| 190 | msb_sixteen_bit_1 | 0.867804824 | 27648 | 108 | 264 | 1 |
| 241 | lsb_sixteen_bit_0 | 0.867771822 | 30078 | 385 | 210 | 1 |
| 241 | msb_fifteen_bit_3 | 0.867569042 | 234 | 118 | 2 | 1 |
| 241 | msb_sixteen_bit_3 | 0.867569042 | 234 | 118 | 2 | 1 |
| 241 | msb_nine_bit_3 | 0.867569042 | 234 | 118 | 2 | 1 |
| 241 | lsb_eleven_bit_1 | 0.867569042 | 117 | 118 | 1 | 1 |
| 241 | msb_ten_bit_3 | 0.867569042 | 234 | 118 | 2 | 1 |
| 241 | lsb_nine_bit_1 | 0.867569042 | 117 | 118 | 1 | 1 |
| 241 | msb_fourteen_bit_3 | 0.867569042 | 234 | 118 | 2 | 1 |
| 241 | msb_twelve_bit_3 | 0.867569042 | 234 | 118 | 2 | 1 |
| 241 | lsb_sixteen_bit_3 | 0.867569042 | 59904 | 118 | 422 | 1 |

| 241 | msb_thirteen_bit_3 | 0.867569042 | 234 | 118 | 2 | 1 |
| --- | --- | --- | --- | --- | --- | --- |
| 241 | lsb_fourteen_bit_1 | 0.867569042 | 117 | 118 | 1 | 1 |
| 241 | lsb_thirteen_bit_1 | 0.867569042 | 117 | 118 | 1 | 1 |
| 241 | msb_nine_bit_0 | 0.867569042 | 117 | 118 | 1 | 1 |
| 241 | lsb_twelve_bit_1 | 0.867569042 | 117 | 118 | 1 | 1 |
| 241 | byte_4 | 0.867569042 | 234 | 118 | 2 | 1 |
| 241 | byte_1 | 0.867569042 | 117 | 118 | 1 | 1 |
| 241 | lsb_ten_bit_1 | 0.867569042 | 117 | 118 | 1 | 1 |
| 241 | msb_eleven_bit_3 | 0.867569042 | 234 | 118 | 2 | 1 |
| 241 | msb_fifteen_bit_1 | 0.867568804 | 29952 | 192 | 216 | 1 |
| 241 | msb_sixteen_bit_1 | 0.867568804 | 29952 | 192 | 216 | 1 |
| 241 | msb_sixteen_bit_4 | 0.867568432 | 59907 | 416 | 434 | 1 |

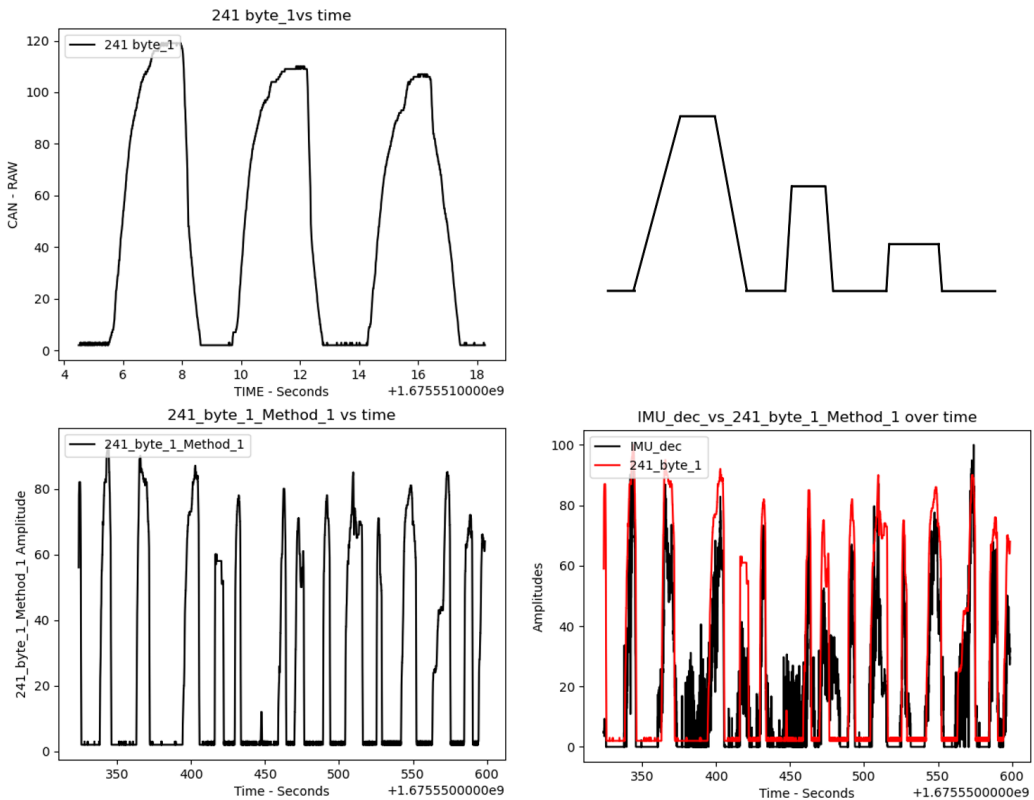

**Figure A5.** 2021 GMC Sierra Brake Pedal Channel 241_byte_1.

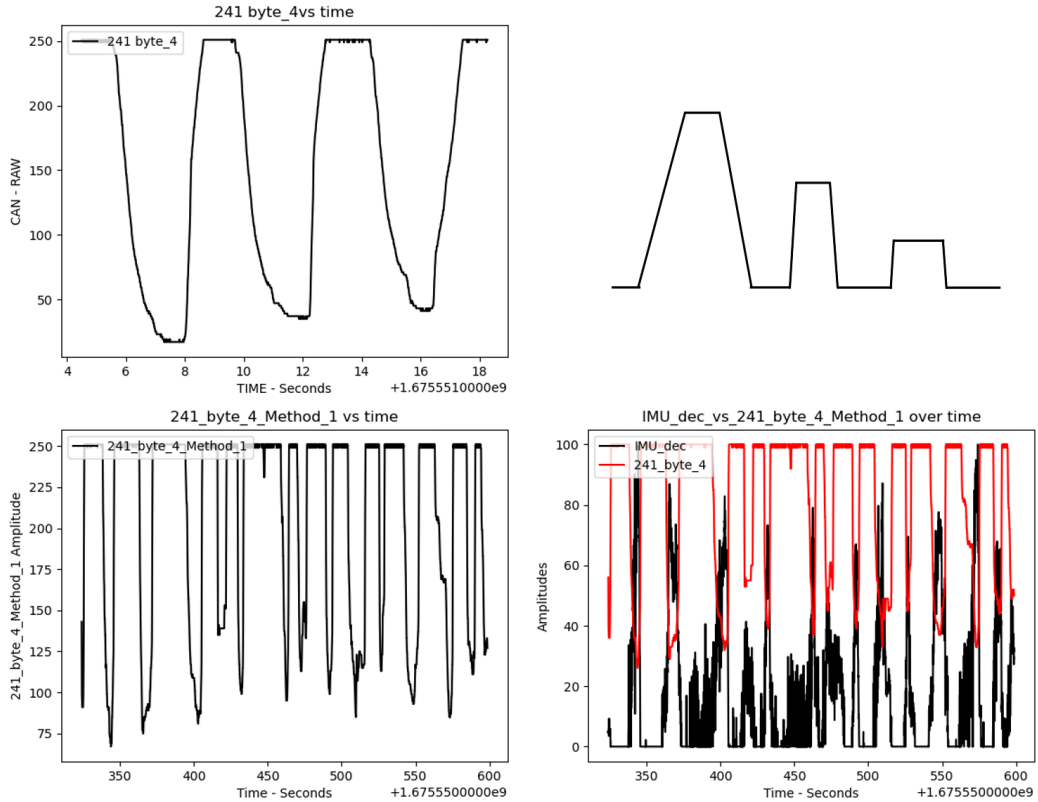

**Figure A6.** 2021 GMC Sierra Brake Pedal Channel 241_byte_4.

*A.3. 2022 Chevrolet Traverse Results*

This section presents data for the 2022 Chevrolet Traverse. Table A3 shows the top 25 potential CAN channels for the accelerator pedal. Figure A7 illustrates the results for CAN channel 190_byte_2 and Figure A8 presents data for CAN channel 401_byte_7. In each figure, the top images show the actual and projected CAN channel data during the accelerator pedal calibration recording. The bottom images show the CAN channel data during the vehicle trip recording.

**Table A3.** 2022 Chevrolet Traverse Accelerator Pedal Results.

| ID | Channel | Correlation | Range | Unique | StDev(*) | Smooth |
|---|---|---|---|---|---|---|
| 401 | lsb_sixteen_bit_5 | 0.847688827 | 65024 | 193 | 726 | 2 |
| 190 | lsb_nine_bit_2 | 0.847320242 | 255 | 192 | 3 | 2 |
| 190 | byte_2 | 0.847320242 | 255 | 192 | 3 | 2 |
| 190 | lsb_eleven_bit_2 | 0.847320242 | 255 | 192 | 3 | 2 |
| 190 | lsb_ten_bit_2 | 0.847320242 | 255 | 192 | 3 | 2 |
| 190 | lsb_thirteen_bit_2 | 0.847320242 | 255 | 192 | 3 | 2 |
| 190 | lsb_twelve_bit_2 | 0.847320242 | 255 | 192 | 3 | 2 |
| 190 | msb_sixteen_bit_2 | 0.847310055 | 65280 | 192 | 727 | 2 |
| 190 | lsb_sixteen_bit_1 | 0.847279596 | 65280 | 192 | 727 | 2 |
| 401 | msb_sixteen_bit_6 | 0.847096049 | 65278 | 193 | 729 | 2 |
| 401 | lsb_sixteen_bit_6 | 0.847096049 | 65278 | 193 | 729 | 2 |

| 401 | lsb_nine_bit_6 | 0.847096049 | 254 | 193 | 3 | 2 |
| 401 | lsb_eleven_bit_6 | 0.847096049 | 254 | 193 | 3 | 2 |
| 401 | lsb_fourteen_bit_6 | 0.847096049 | 254 | 193 | 3 | 2 |
| 401 | lsb_twelve_bit_6 | 0.847096049 | 254 | 193 | 3 | 2 |
| 401 | byte_7 | 0.847096049 | 254 | 193 | 3 | 2 |
| 401 | lsb_thirteen_bit_6 | 0.847096049 | 254 | 193 | 3 | 2 |
| 401 | byte_6 | 0.847096049 | 254 | 193 | 3 | 2 |
| 401 | lsb_fifteen_bit_6 | 0.847096049 | 254 | 193 | 3 | 2 |
| 401 | lsb_ten_bit_6 | 0.847096049 | 254 | 193 | 3 | 2 |
| 201 | lsb_eleven_bit_4 | 0.847081921 | 254 | 194 | 3 | 2 |
| 201 | lsb_fourteen_bit_4 | 0.847081921 | 254 | 194 | 3 | 2 |
| 201 | byte_4 | 0.847081921 | 254 | 194 | 3 | 2 |
| 201 | lsb_ten_bit_4 | 0.847081921 | 254 | 194 | 3 | 2 |
| 201 | lsb_nine_bit_4 | 0.847081921 | 254 | 194 | 3 | 2 |

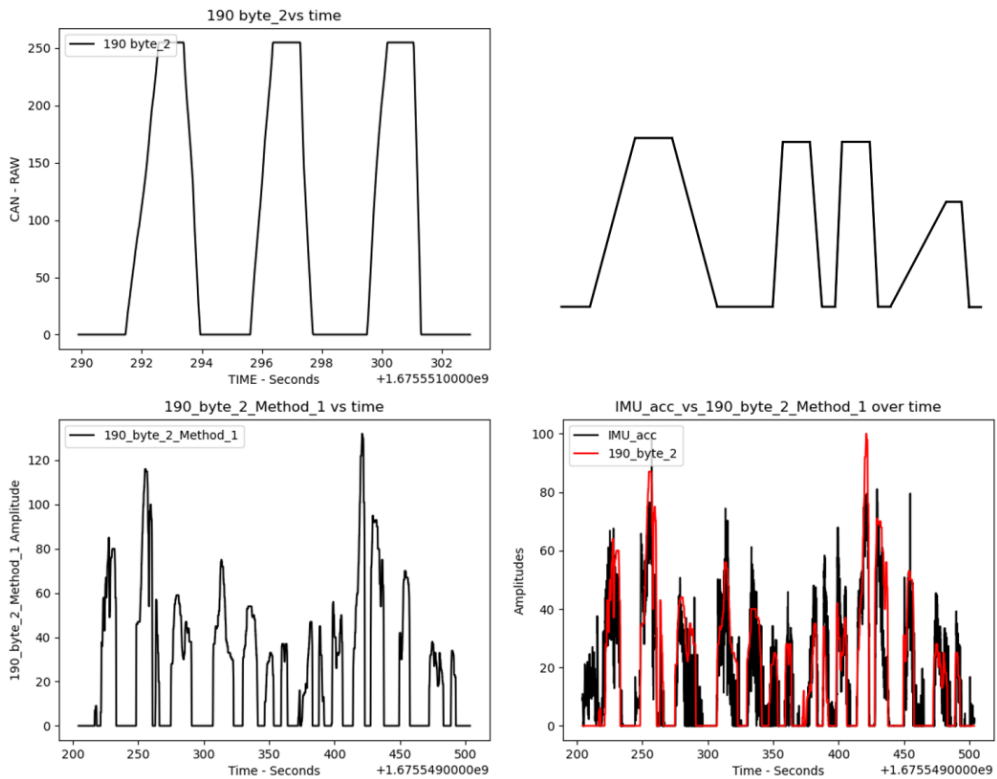

**Figure A7.** 2022 Chevrolet Traverse Accelerator Pedal Channel 190_byte_2.

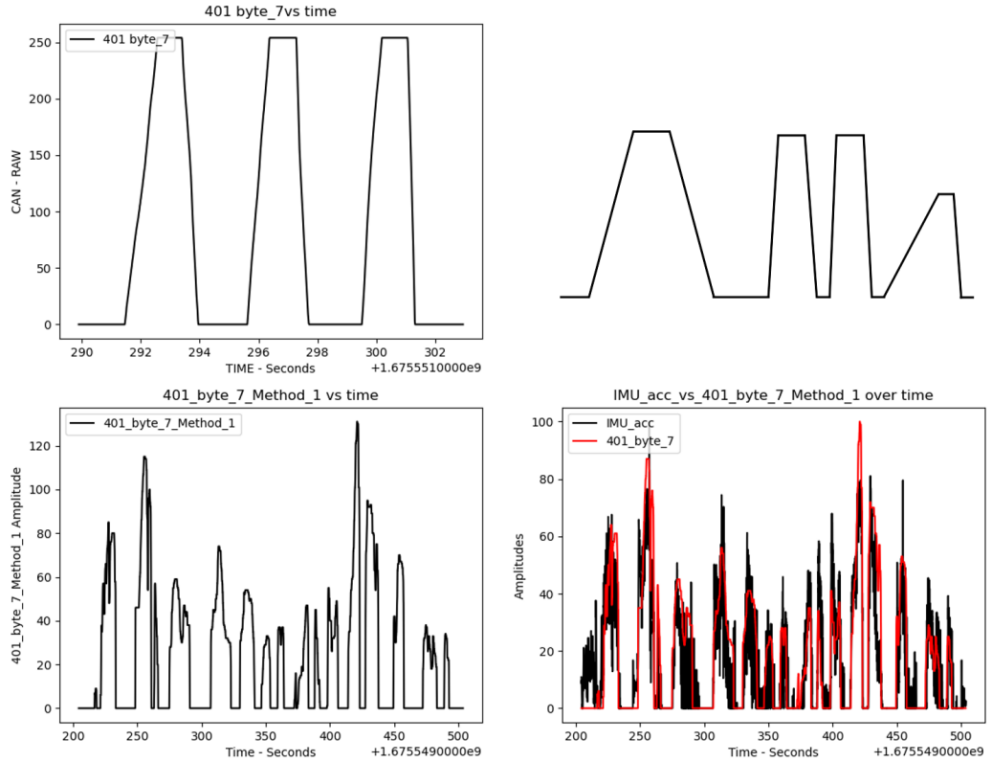

**Figure A8.** 2022 Chevrolet Traverse Accelerator Pedal Channel 401_byte_7.

Table A4 lists the potential CAN channels for the brake pedal. Figure A9 presents data for CAN channel 241_byte_1 and Figure A10 presents data for CAN channel 190_msb_sixteen_bit_1. For each figure, the top images show the actual and projected CAN channel data during the brake pedal calibration recording. The bottom images show the CAN channel data during the vehicle trip recording.

**Table A4.** 2022 Chevrolet Traverse Brake Pedal Results.

| ID | Channel | Correlation | Range | Unique | StDev(*) | Smooth |
|---|---|---|---|---|---|---|
| 241 | lsb_sixteen_bit_0 | 0.829625448 | 34942 | 315 | 219 | 1 |
| 241 | msb_fifteen_bit_3 | 0.82916633 | 272 | 135 | 2 | 1 |
| 241 | msb_sixteen_bit_3 | 0.82916633 | 272 | 135 | 2 | 1 |
| 241 | msb_nine_bit_3 | 0.82916633 | 272 | 135 | 2 | 1 |
| 241 | msb_twelve_bit_3 | 0.82916633 | 272 | 135 | 2 | 1 |
| 241 | lsb_twelve_bit_1 | 0.82916633 | 136 | 135 | 1 | 1 |
| 241 | msb_thirteen_bit_3 | 0.82916633 | 272 | 135 | 2 | 1 |
| 241 | lsb_eleven_bit_1 | 0.82916633 | 136 | 135 | 1 | 1 |
| 241 | msb_ten_bit_3 | 0.82916633 | 272 | 135 | 2 | 1 |
| 241 | lsb_nine_bit_1 | 0.82916633 | 136 | 135 | 1 | 1 |
| 241 | lsb_fourteen_bit_1 | 0.82916633 | 136 | 135 | 1 | 1 |
| 241 | msb_nine_bit_0 | 0.82916633 | 136 | 135 | 1 | 1 |
| 241 | lsb_ten_bit_1 | 0.82916633 | 136 | 135 | 1 | 1 |
| 241 | msb_fourteen_bit_3 | 0.82916633 | 272 | 135 | 2 | 1 |

| 241 | msb_eleven_bit_3 | 0.82916633 | 272 | 135 | 2 | 1 |
| --- | --- | --- | --- | --- | --- | --- |
| 241 | lsb_thirteen_bit_1 | 0.82916633 | 136 | 135 | 1 | 1 |
| 241 | byte_1 | 0.82916633 | 136 | 135 | 1 | 1 |
| 241 | msb_sixteen_bit_1 | 0.829165875 | 34816 | 189 | 260 | 1 |
| 190 | lsb_sixteen_bit_0 | 0.826182161 | 34316 | 263 | 275 | 1 |
| 190 | msb_sixteen_bit_1 | 0.826180292 | 34304 | 127 | 327 | 1 |
| 844 | lsb_sixteen_bit_3 | 0.592236228 | 255 | 4 | 1 | 1 |
| 241 | lsb_nine_bit_3 | 0.517030715 | 128 | 3 | 1 | 1 |
| 241 | lsb_ten_bit_3 | 0.517030715 | 128 | 3 | 1 | 1 |

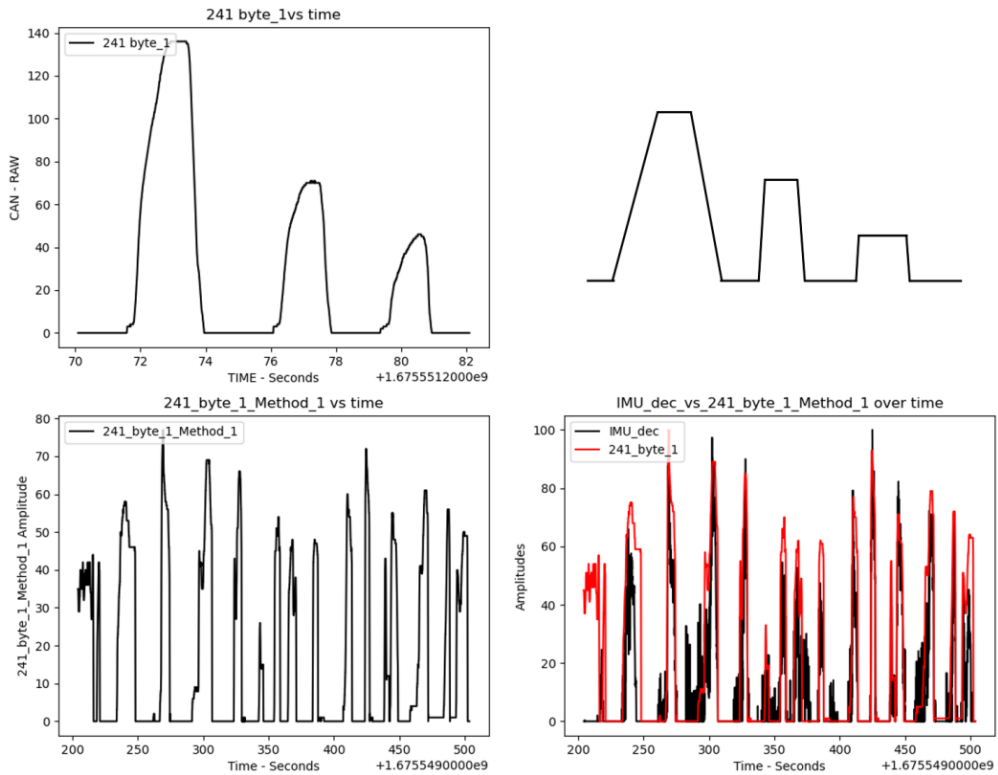

**Figure A9.** 2022 Chevrolet Traverse Brake Pedal Channel 241_byte_1.

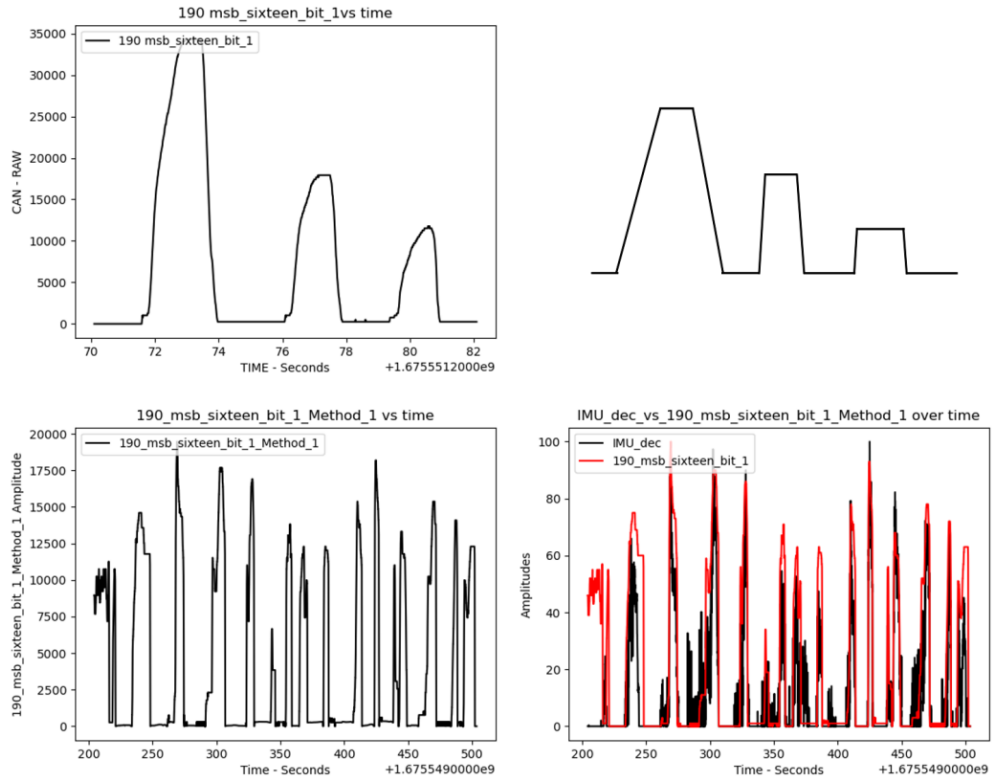

**Figure A10.** 2022 Chevrolet Traverse Brake Pedal Channel 190_msb_sixteen_bit_1.

*A.4. 2009 Chevrolet Impala Results*

This section presents data for the 2009 Chevrolet Impala. Table A5 lists the potential CAN channels for the accelerator pedal. Figure A11 presents data for CAN channel 201_byte_4 and Figure A12 presents data for CAN channel 201_msb_sixteen_bit_4. For each figure, the top images show the actual and projected CAN channel data during the accelerator pedal calibration recording. The bottom images show the CAN channel data during the vehicle trip recording.

**Table A5.** 2009 Chevrolet Impala Accelerator Pedal Results.

| ID | Channel | Correlation | Range | Unique | StDev(*) | Smooth |
|---|---|---|---|---|---|---|
| 201 | lsb_nine_bit_4 | 0.873412116 | 254 | 160 | 4 | 2 |
| 201 | byte_4 | 0.873412116 | 254 | 160 | 4 | 2 |
| 201 | msb_sixteen_bit_4 | 0.873408622 | 65024 | 160 | 928 | 2 |
| 201 | lsb_sixteen_bit_3 | 0.873394438 | 65052 | 219 | 763 | 2 |

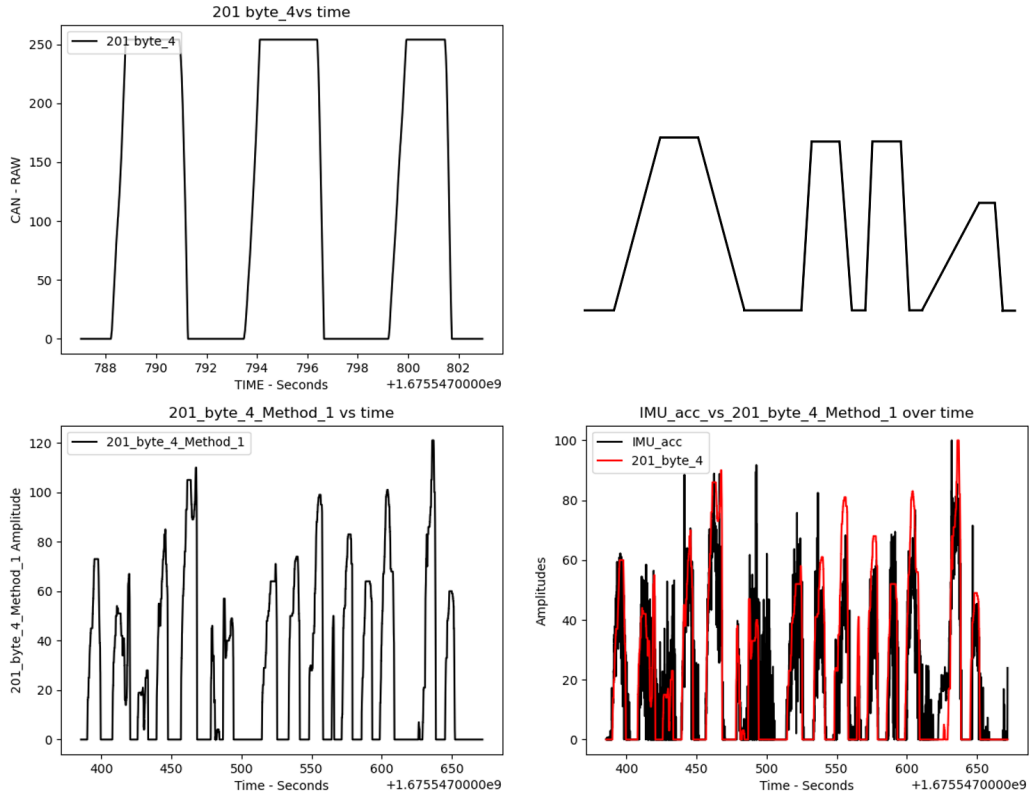

**Figure A11.** 2009 Chevrolet Impala Accelerator Pedal Channel 201_byte_4.

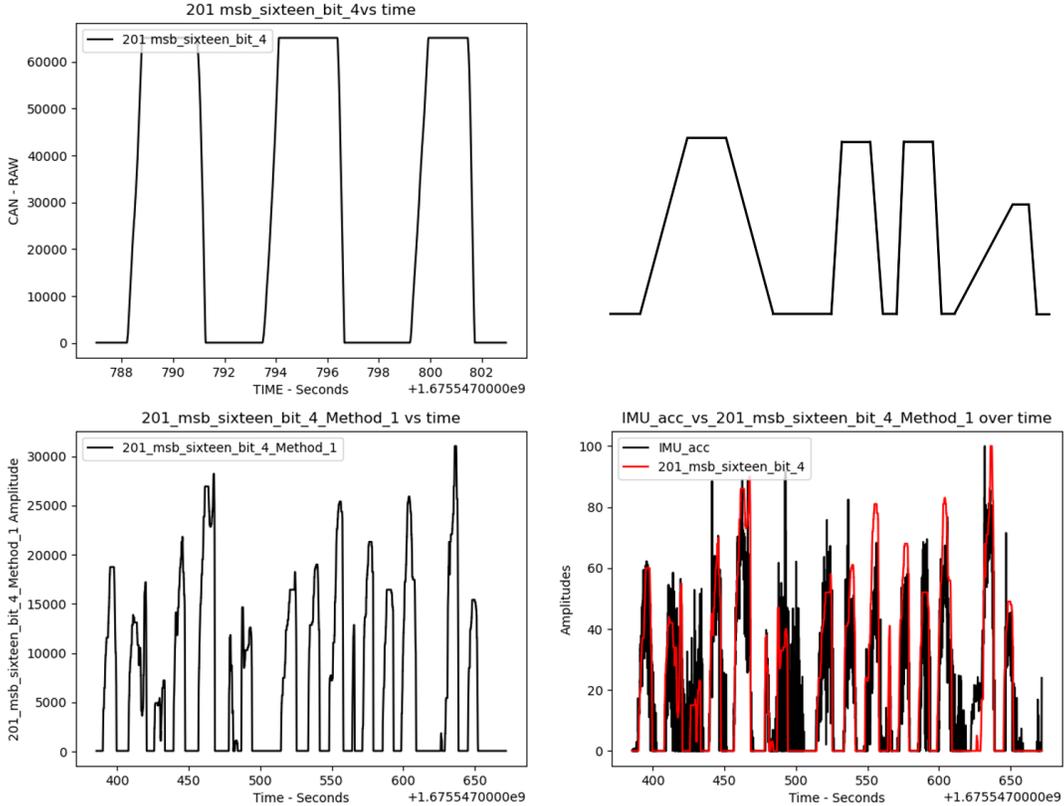

**Figure A12.** 2009 Chevrolet Impala Accelerator Pedal Channel 201_msb_sixteen_bit_4.

Table A6 lists the potential CAN channels for the brake pedal. Figure A13 presents data for CAN channel 241_msb_sixteen_bit_1 and A14 presents data for CAN channel 241_lsb_sixteen_bit_0. For each figure, the top images show the actual and projected CAN channel data during the brake pedal calibration recording. The bottom images show the CAN channel data during the vehicle trip recording.

**Table A6.** 2009 Chevrolet Impala Brake Pedal Results.

| ID | Channel | Correlation | Range | Unique | StDev(*) | Smooth |
|---|---|---|---|---|---|---|
| 241 | lsb_sixteen_bit_0 | 0.847027986 | 33662 | 412 | 224 | 1 |
| 241 | msb_sixteen_bit_1 | 0.846750341 | 33536 | 278 | 265 | 1 |

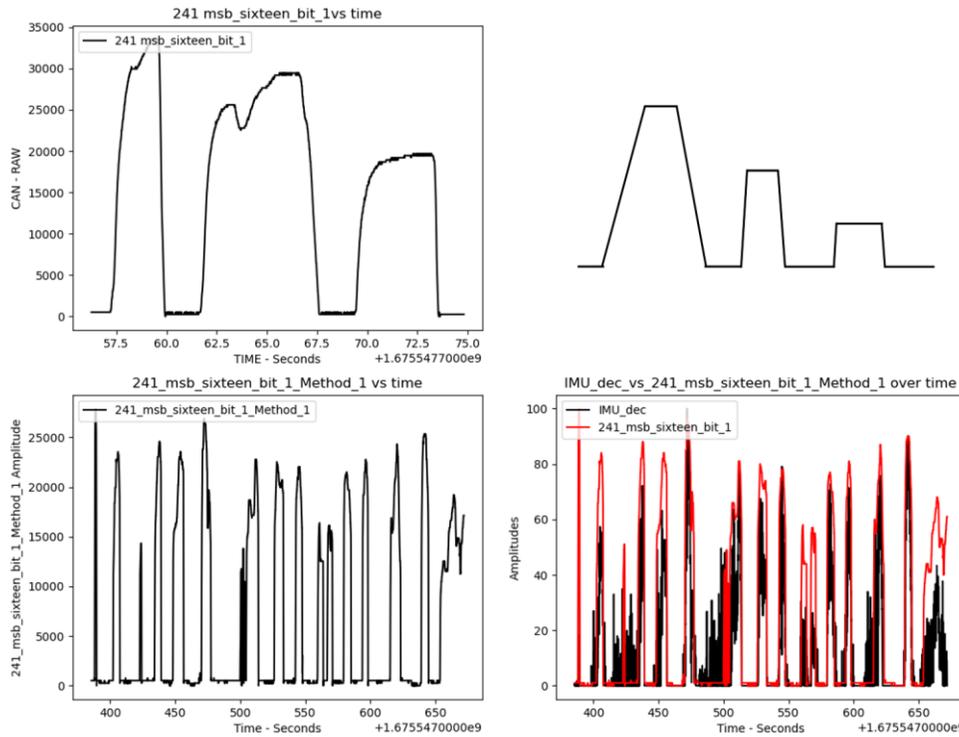

**Figure A13.** 2009 Chevrolet Impala Brake Pedal Channel 241_msb_sixteen_bit_1.

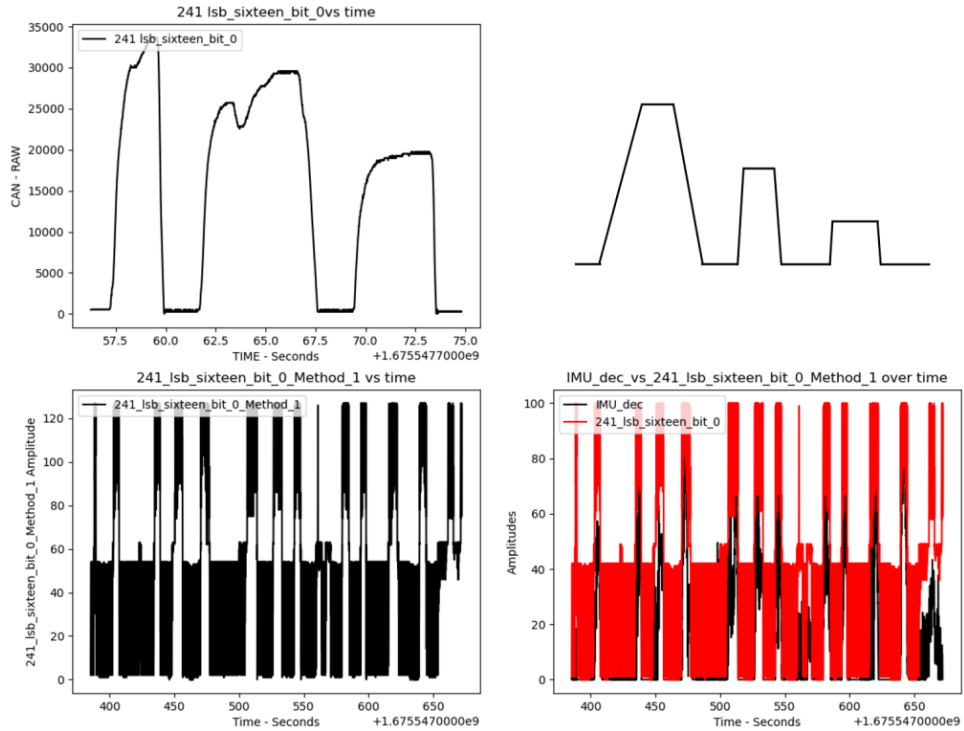

**Figure A14.** 2009 Chevrolet Impala Brake Pedal Channel 241_lsb_sixteen_bit_0.

*A.5. 2006 Volvo XC90 Results*

This section presents data for the 2006 Volvo XC90. Table A7 lists the potential CAN channels for the accelerator pedal. Figure A15 presents data for CAN channel 6438942_msb_sixteen_bit_6 and Figure A16 presents data for CAN channel 10616854_msb_fourteen_bit_2. For each figure, the top images show the actual and projected CAN channel data from during the accelerator pedal calibration recording. The bottom images show the CAN channel data during the vehicle trip recording.

**Table A7.** 2006 Volvo XC90 Accelerator Pedal Results.

| ID | Channel | Correlation | Range | Unique | StDev(*) | Smooth |
|---|---|---|---|---|---|---|
| 10616854 | msb_thirteen_bit_2 | 0.797618521 | 999 | 284 | 9 | 1 |
| 10616854 | msb_fourteen_bit_2 | 0.797618521 | 999 | 284 | 9 | 1 |
| 10616854 | msb_ten_bit_2 | 0.797618521 | 999 | 284 | 9 | 1 |
| 10616854 | msb_twelve_bit_2 | 0.797618521 | 999 | 284 | 9 | 1 |
| 10616854 | msb_eleven_bit_2 | 0.797618521 | 999 | 284 | 9 | 1 |
| 6438942 | msb_sixteen_bit_6 | 0.797201935 | 999 | 282 | 9 | 1 |
| 6438942 | msb_fourteen_bit_6 | 0.797201935 | 999 | 282 | 9 | 1 |
| 6438942 | msb_fifteen_bit_6 | 0.797201935 | 999 | 282 | 9 | 1 |
| 6438942 | msb_ten_bit_6 | 0.797201935 | 999 | 282 | 9 | 1 |
| 6438942 | msb_eleven_bit_6 | 0.797201935 | 999 | 282 | 9 | 1 |
| 6438942 | msb_thirteen_bit_6 | 0.797201935 | 999 | 282 | 9 | 1 |
| 6438942 | msb_twelve_bit_6 | 0.797201935 | 999 | 282 | 9 | 1 |

| 6438942 | lsb_nine_bit_6 | 0.634241021 | 131 | 8 | 1 | 1 |

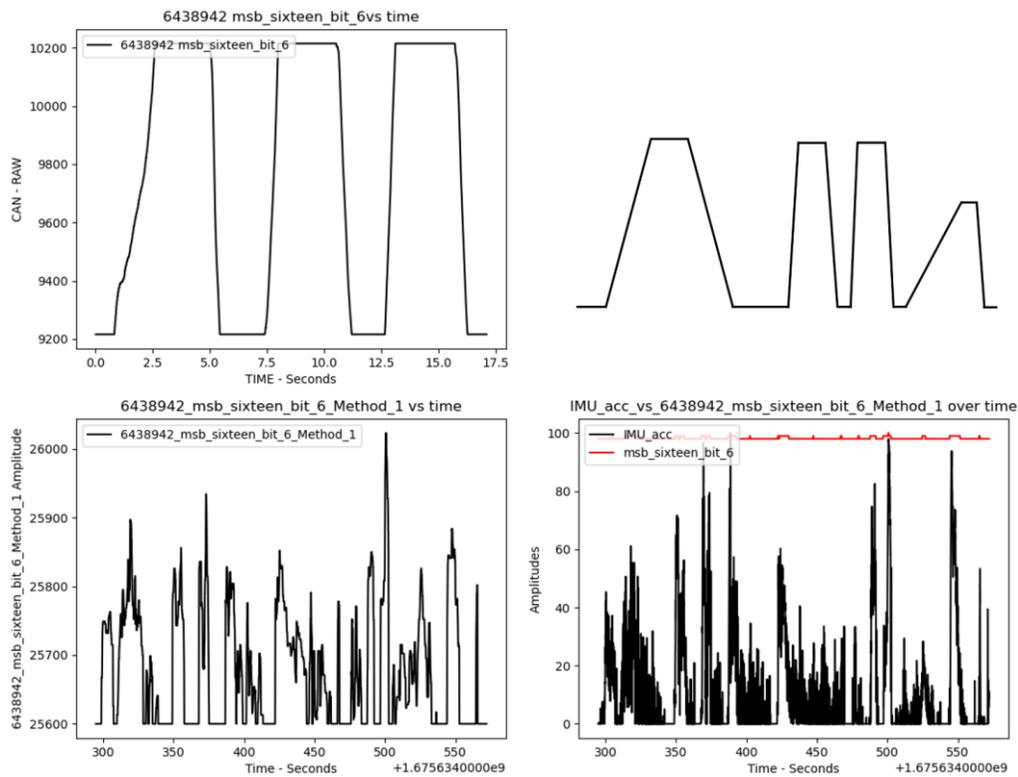

**Figure A15.** 2006 Volvo XC90 Accelerator Pedal Channel 6438942_msb_sixteen_bit_6.

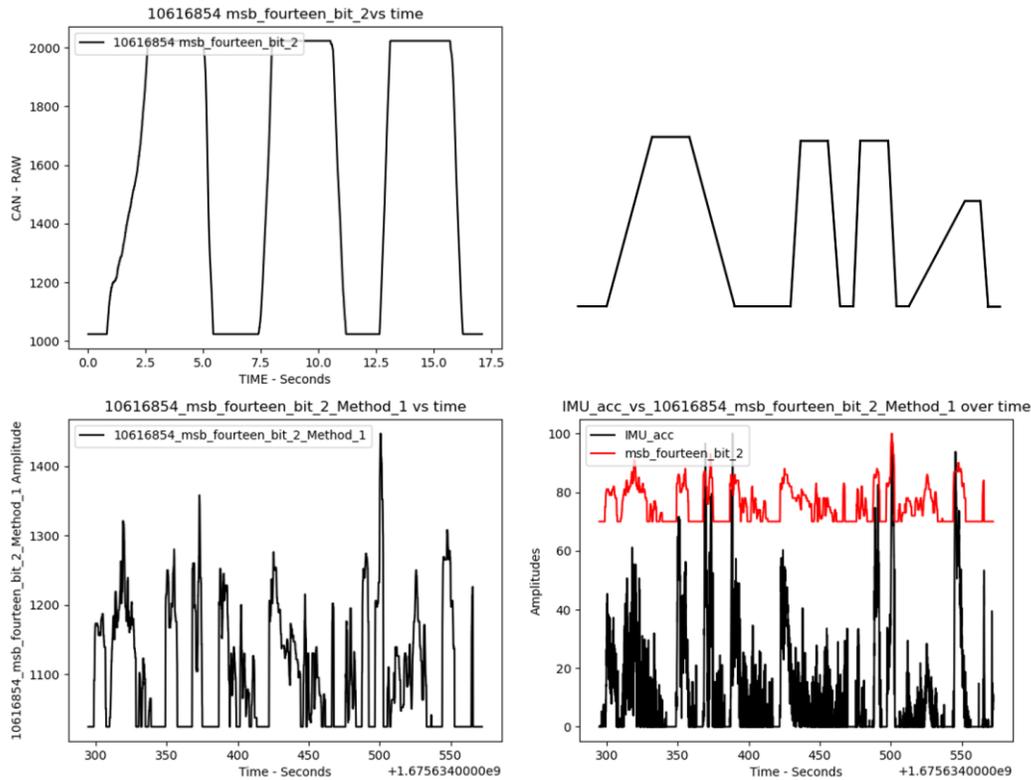
**Figure A16.** 2006 Volvo XC90 Accelerator Pedal Channel 10616854_msb_fourteen_bit_2.

Table A8 lists the potential CAN channels for the brake pedal. Figure A17 shows data for CAN channel 2244644_msb_sixteen_bit_2 and Figure A18 presents data for CAN channel 2244644_msb_ten_bit_6. For each figure, the top images show the actual and projected CAN channel data from during the brake pedal calibration recording. The bottom images show the CAN channel data during the vehicle trip recording.

**Table A8.** 2006 Volvo XC90 Brake Pedal Results.

| ID | Channel | Correlation | Range | Unique | StDev(*) | Smooth |
|---|---|---|---|---|---|---|
| 2244644 | msb_sixteen_bit_2 | 0.818554642 | 25088 | 99 | 164 | 1 |
| 2244644 | msb_fifteen_bit_2 | 0.818554642 | 25088 | 99 | 164 | 1 |
| 2244644 | lsb_sixteen_bit_1 | 0.818485393 | 25088 | 99 | 164 | 1 |
| 2244644 | msb_ten_bit_6 | 0.724439439 | 740 | 353 | 6 | 1 |
| 2244644 | msb_twelve_bit_6 | 0.724439439 | 740 | 353 | 6 | 1 |
| 2244644 | msb_eleven_bit_6 | 0.724439439 | 740 | 353 | 6 | 1 |

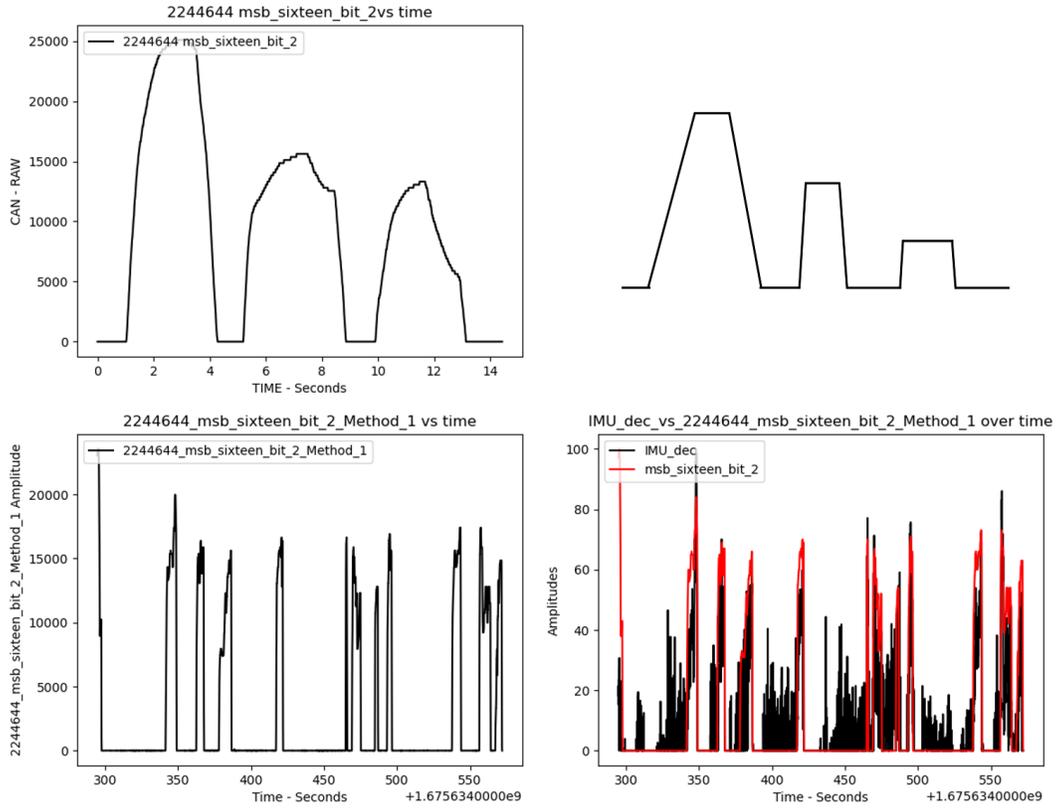

**Figure A17.** 2006 Volvo XC90 Brake Pedal Channel 2244644_msb_sixteen_bit_2.

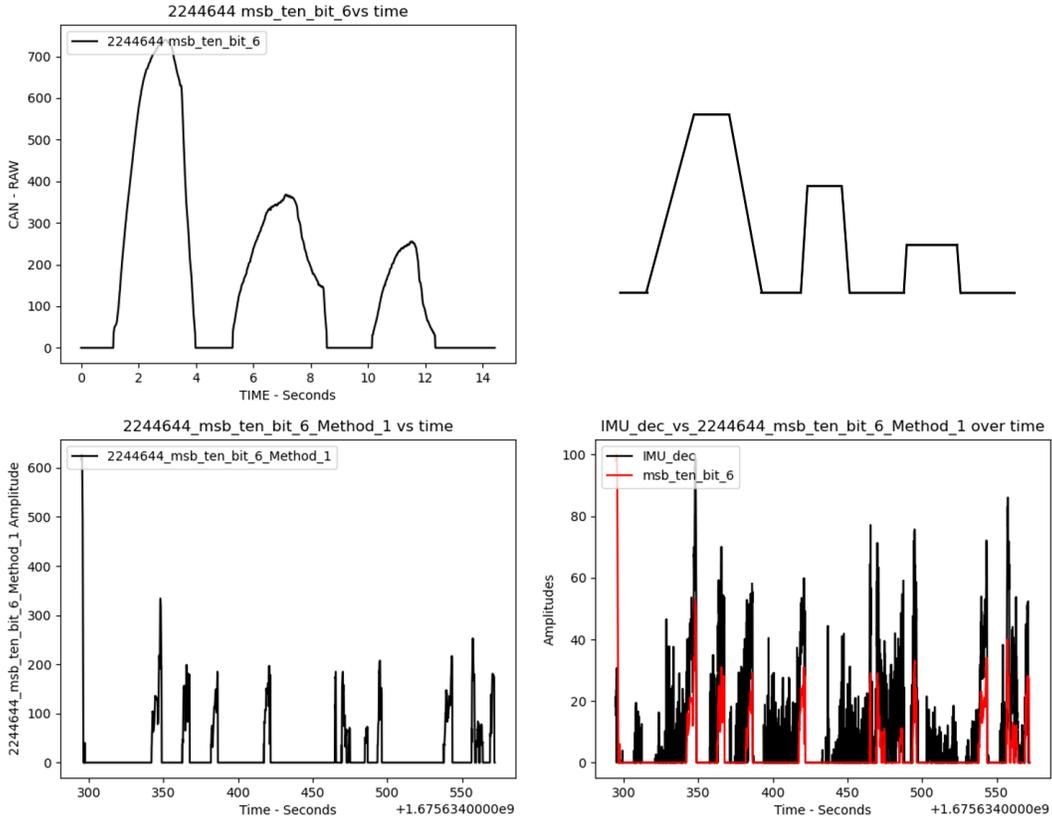

**Figure A18.** 2006 Volvo XC90 Brake Pedal Channel 2244644_msb_ten_bit_6.

*A.6. 2016 Ford Fusion Results*

This section presents data for the 2016 Ford Fusion. Table A9 lists the potential CAN channels for the accelerator pedal. Figure A19 presents data for CAN channel 516_msb_ten_bit_0 and Figure A20 presents data for CAN channel 516_msb_sixteen_bit_0. For each figure, the top images show the actual and projected CAN channel data from during the accelerator pedal calibration recording. The bottom images show the CAN channel data during the vehicle trip recording.

**Table A9.** 2016 Ford Fusion Accelerator Pedal Results.

| ID | Channel | Correlation | Range | Unique | StDev(*) | Smooth |
|---|---|---|---|---|---|---|
| 516 | msb_fifteen_bit_0 | 0.882773625 | 999 | 95 | 29 | 3 |
| 516 | msb_sixteen_bit_0 | 0.882773625 | 999 | 95 | 29 | 3 |
| 516 | msb_ten_bit_0 | 0.882773625 | 999 | 95 | 29 | 3 |
| 516 | msb_eleven_bit_0 | 0.882773625 | 999 | 95 | 29 | 3 |
| 516 | msb_twelve_bit_0 | 0.882773625 | 999 | 95 | 29 | 3 |
| 516 | msb_thirteen_bit_0 | 0.882773625 | 999 | 95 | 29 | 3 |
| 516 | msb_fourteen_bit_0 | 0.882773625 | 999 | 95 | 29 | 3 |

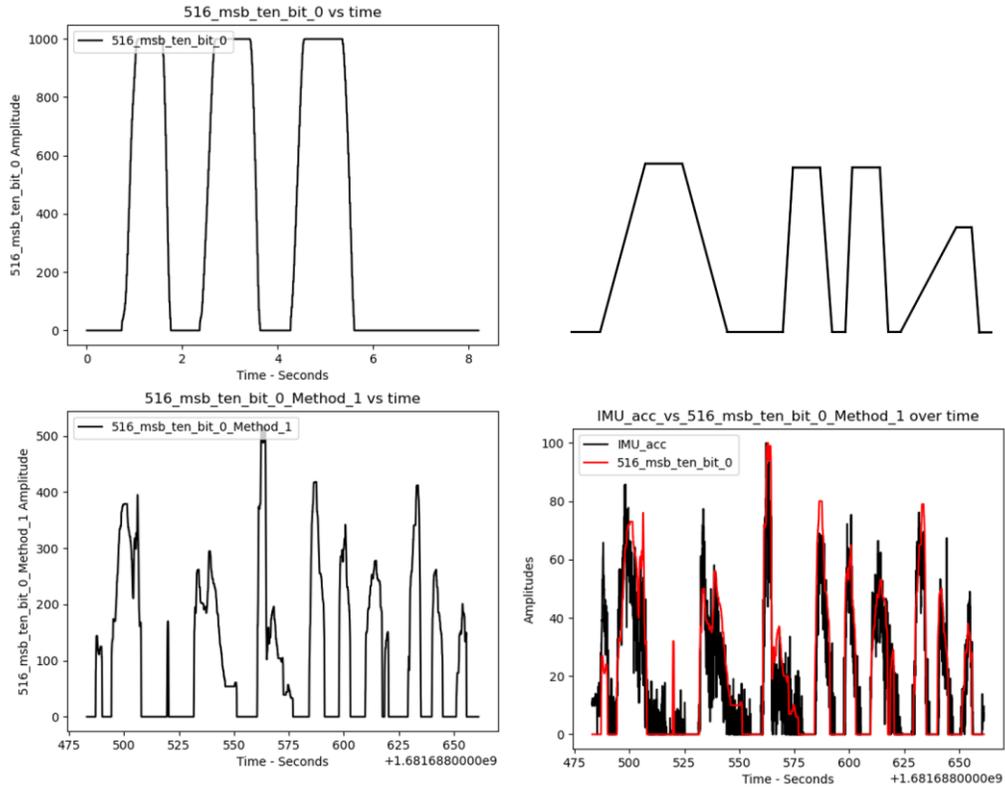

**Figure A19.** 2016 Ford Fusion Accelerator Pedal Channel 516_msb_ten_bit_0.

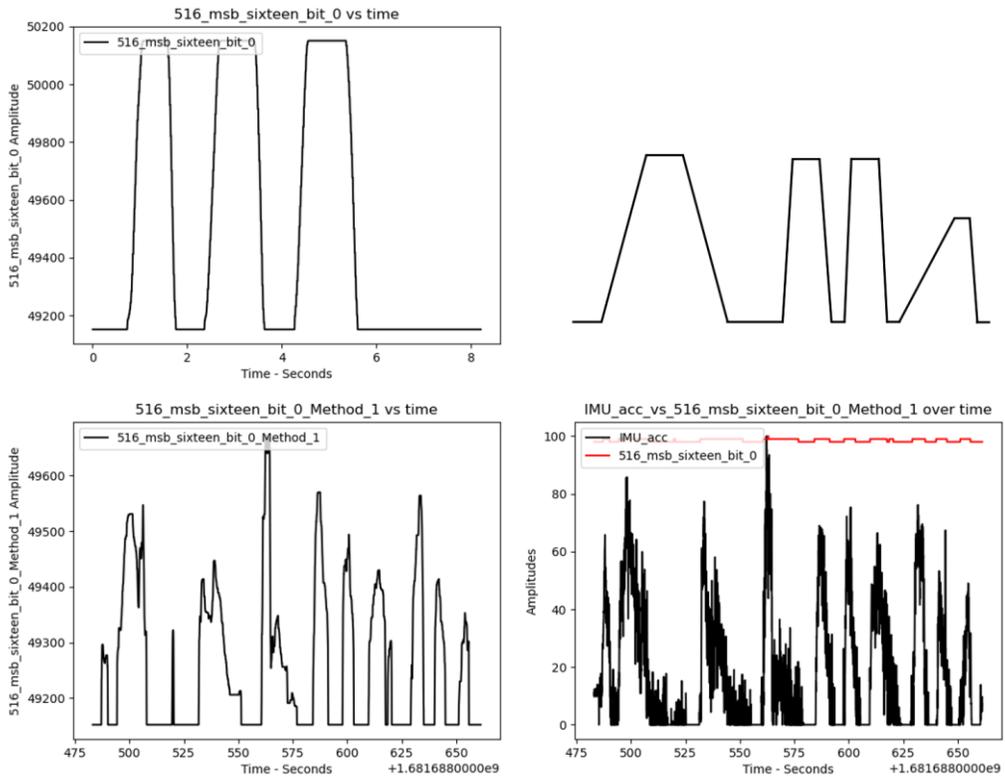

**Figure A20.** 2016 Ford Fusion Accelerator Pedal Channel 516_msb_sixteen_bit_0.

Table A10 lists the potential CAN channels for the brake pedal. Figure A21 presents data for CAN channel 125_msb_ten_bit_3 and Figure A22 presents data for CAN channel 125_msb_twelve_bit_3. For each figure, the top images show the actual and projected CAN channel data from during the brake pedal calibration recording. The bottom images show the CAN channel data during the vehicle trip recording.

**Table A10.** 2016 Ford Fusion Brake Pedal Results.

| ID | Channel | Correlation | Range | Unique | StDev(*) | Smooth |
|---|---|---|---|---|---|---|
| 125 | msb_twelve_bit_3 | 0.804368197 | 645 | 223 | 12 | 2 |
| 125 | msb_eleven_bit_3 | 0.804368197 | 645 | 223 | 12 | 2 |
| 125 | msb_ten_bit_3 | 0.804368197 | 645 | 223 | 12 | 2 |

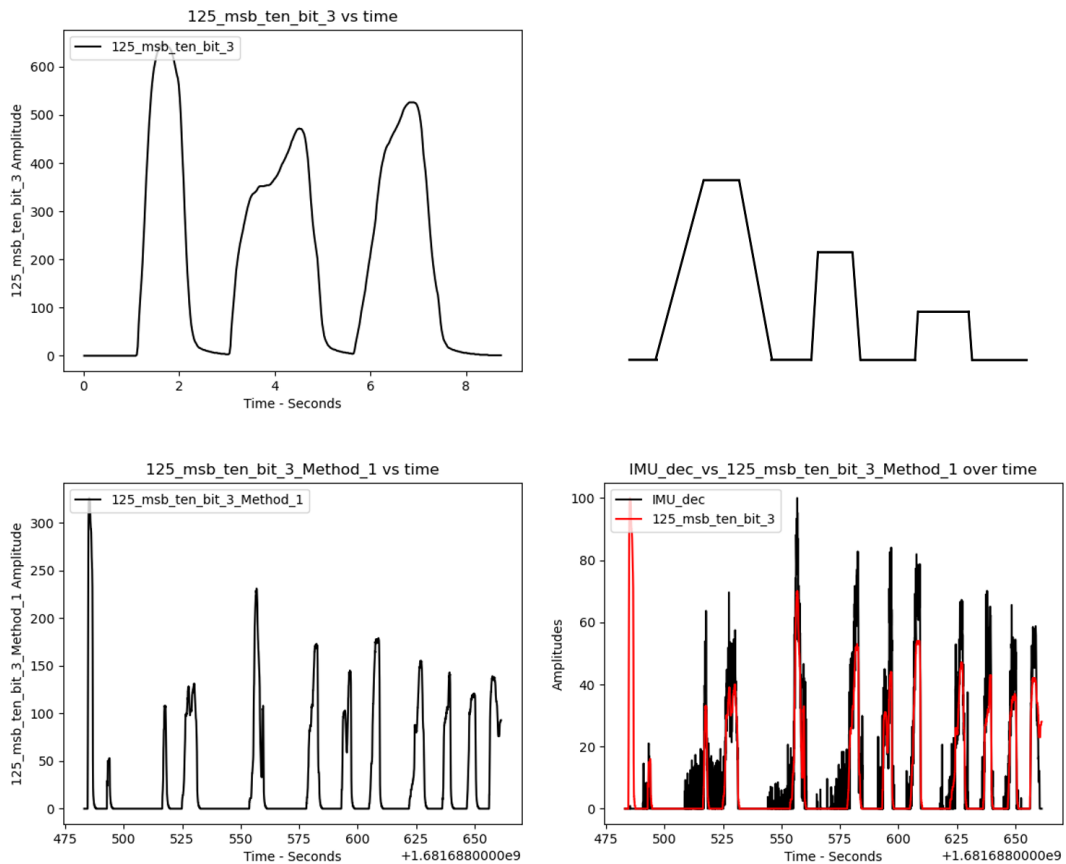

**Figure A21.** 2016 Ford Fusion brake pedal position channel 125_msb_ten_bit_3.

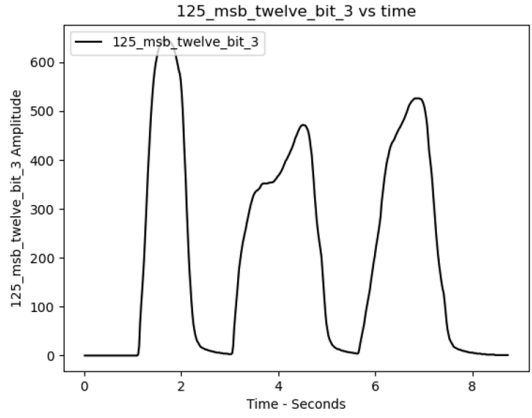
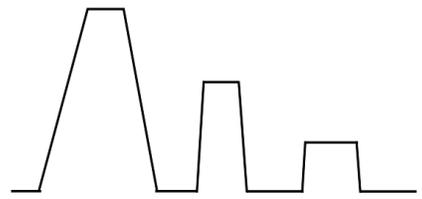
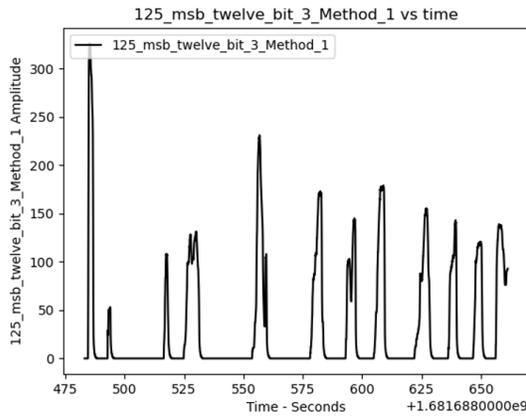
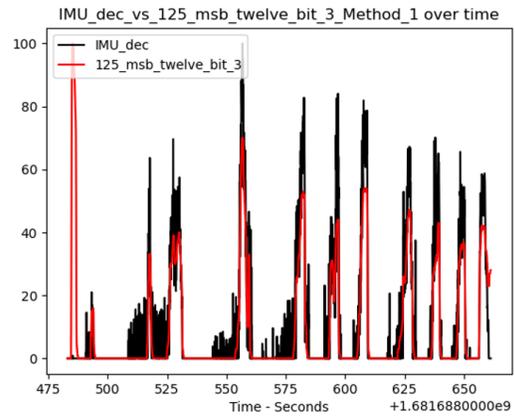

**Figure A22.** 2016 Ford Fusion brake pedal position channel 125_msb_twelve_bit_3.